\documentclass[sigconf]{acmart}

\AtBeginDocument{%
  }

\usepackage{bbding}
\usepackage{subfigure}
\usepackage[boxed,ruled,lined]{algorithm2e}
\usepackage{multirow}
\usepackage{makecell}

\usepackage{arydshln} 
\usepackage{enumerate}
\usepackage{enumitem}
\usepackage{makecell}

\newcommand{\paratitle}[1]{\vspace{0.8ex}\noindent\textbf{#1}}
\newcommand{\ourname}{CFRAG\xspace}

\copyrightyear{2025}
\acmYear{2025}
\setcopyright{acmlicensed}
\acmConference[SIGIR '25] {Proceedings of the 48th International ACM SIGIR Conference on Research and Development in Information Retrieval}{ July 13--18, 2025}{Padua, Italy.}
\acmBooktitle{Proceedings of the 48th International ACM SIGIR Conference on Research and Development in Information Retrieval (SIGIR '25), July 13--18, 2025, Padua, Italy}
\acmISBN{979-8-4007-1592-1/25/07}
\acmDOI{10.1145/XXXXXX.XXXXXX}

\begin{document}

\title{Retrieval Augmented Generation with Collaborative Filtering for Personalized Text Generation}

\author{Teng Shi}
\affiliation{%
\institution{Renmin University of China}
  \city{Beijing}\country{China}
  }
\email{shiteng@ruc.edu.cn}

\author{Jun Xu}
\authornote{Corresponding authors. Work partially done at Engineering Research Center of Next-Generation Intelligent Search and Recommendation, Ministry of Education. \\
Work done when Teng Shi was the intern at Kuaishou.
}
\author{Xiao Zhang}
\affiliation{
  \institution{Renmin University of China}
  \city{Beijing}\country{China}
  }
\email{{junxu,zhangx89}@ruc.edu.cn}

\author{Xiaoxue Zang}
\author{Kai Zheng}
\affiliation{
  \institution{Kuaishou Technology Co., Ltd.}
  \city{Beijing}\country{China}
  }
\email{xxic666@126.com}
\email{zhengk92@gmail.com}

\author{Yang Song}
\author{Han Li}
\affiliation{
  \institution{Kuaishou Technology Co., Ltd.}
  \city{Beijing}\country{China}
  }
\email{ys@sonyis.me}
\email{lihan08@kuaishou.com}

\renewcommand{\shortauthors}{Teng Shi et al.}

\begin{abstract}

Recently, the personalization of Large Language Models (LLMs) to generate content that aligns with individual user preferences has garnered widespread attention. Personalized Retrieval-Augmented Generation (RAG), which retrieves relevant documents from the user's history to reflect their preferences and enhance LLM generation, is one commonly used approach for personalization. However, existing personalized RAG methods do not consider that the histories of similar users can also assist in personalized generation for the current user, meaning that collaborative information between users can also benefit personalized generation.
Inspired by the application of collaborative filtering in recommender systems, we propose a method called \textbf{\ourname}, which adapts \textbf{C}ollaborative \textbf{F}iltering to \textbf{RAG} for personalized text generation. However, this presents two challenges: 
(1)~how to incorporate collaborative information without explicit user similarity labels?
(2)~how to retrieve documents that support personalized LLM generation?
For Challenge 1, we use contrastive learning to train user embeddings to retrieve similar users and introduce collaborative information.
For Challenge 2,
we design a personalized retriever and reranker to retrieve the top-$k$ documents from these users' histories. We take into account the user's preference during retrieval and reranking. Then we leverage feedback from the LLM to fine-tune the personalized retriever and reranker, enabling them to retrieve documents that meet the personalized generation needs of the LLM.
Experimental results on the Language Model Personalization (LaMP) benchmark validate the effectiveness of \ourname. 
Further analysis confirms the importance of incorporating collaborative information.

\end{abstract}




\maketitle

\section{Introduction}
\label{sec:intro}

Personalizing Large Language Models (LLMs)~\cite{zhao2023survey} to generate personalized outputs tailored to individual user preferences has emerged as a significant and rapidly growing field~\cite{richardson2023integrating,jang2023personalized,salemi2023lamp,salemi2024optimization,zhuang2024hydra,li2024personalized,tan2024personalized,tan2024democratizinglargelanguagemodels}.
Personalized Retrieval-Augmented Generation (RAG)~\cite{gao2023retrieval} has become a commonly used approach for personalizing LLMs~\cite{richardson2023integrating,salemi2023lamp,salemi2024optimization,zhuang2024hydra}.

\begin{figure*}[t]
    \centering
    \includegraphics[width=0.95\linewidth]{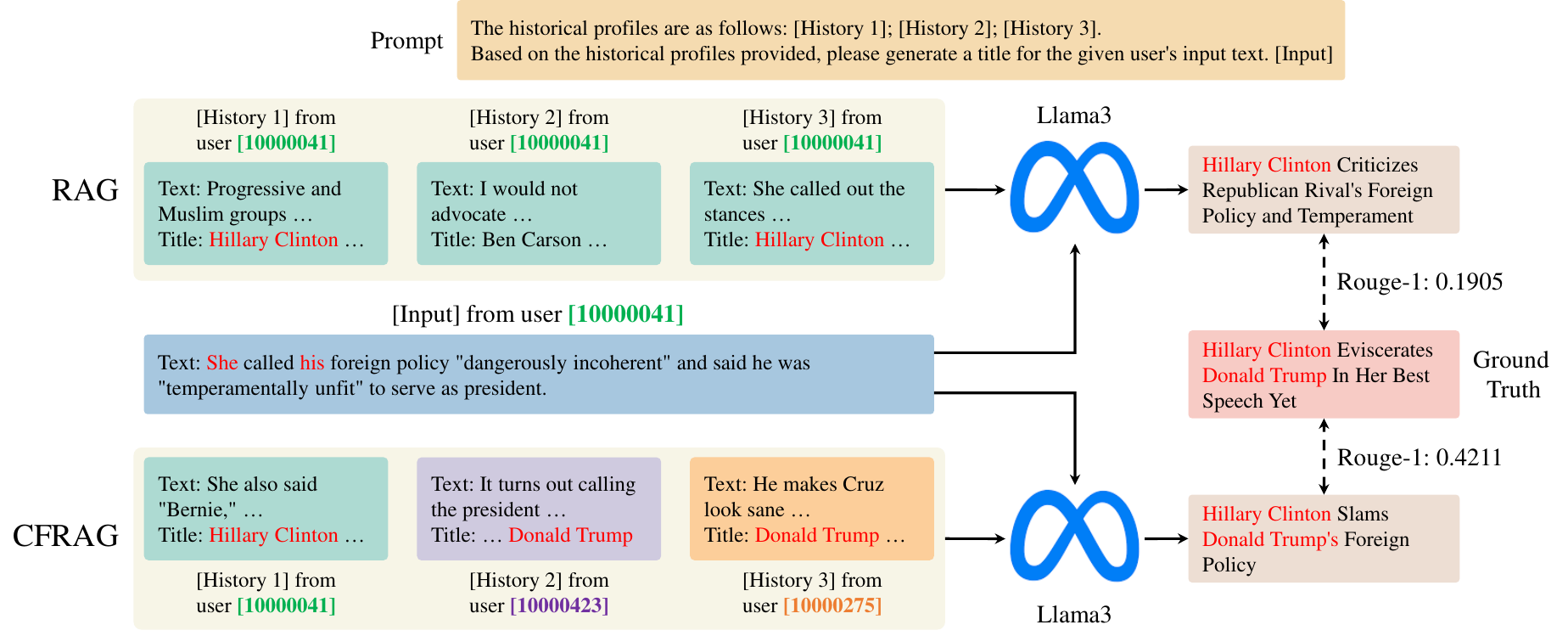}
   \vspace{-10px}
    \caption{
    An example from the LaMP-4 dataset~\cite{salemi2023lamp}. The task of LaMP-4 is to generate personalized news headlines based on user input. This example illustrates the benefit of collaborative information for LLM personalization: (a) The top shows results retrieved by the existing RAG method from the current user's history, where we can only infer that ``She'' in the user's input refers to ``Hillary Clinton'‘. (b) The bottom shows results retrieved by our method from similar users' histories, allowing us to infer further that ``his'' in the user's input refers to ``Donald Trump'' thus enabling the generation of a more accurate result.
    }
    \label{fig:introduction-example}
   \vspace{-0.3cm}
\end{figure*}

The process of existing personalized RAG methods typically involves retrieving similar documents from the user's historical behaviors based on the user's input query, then concatenating these documents with the query as a prompt input to the LLM for generation.
Although effective, this approach is limited to retrieving only the current user's history, neglecting collaborative information. Users with similar histories tend to be more alike, and the information from these similar users can also aid in personalizing generation for the current user.
As shown in the example in Figure~\ref{fig:introduction-example}, the upper part illustrates the results of the existing RAG method, which retrieves documents from the current user's history. We can only infer from these results that ``She'' in the user's input refers to ``Hillary Clinton''. In contrast, the lower part demonstrates our method, which retrieves documents from the history of similar users. In this case, we can further infer that ``his'' in the user's input refers to ``Donald Trump'', leading to a better generation result.
From this example, we can see that incorporating collaborative information allows the retrieval of more diverse documents, helping the LLM generate results that better meet the user's needs.

Inspired by the application of collaborative filtering in recommender systems~\cite{xue2017deep,he2017neural,wang2019neural}, we propose to adapt collaborative information into RAG to personalize LLMs. 
However, adapting collaborative filtering to personalized RAG presents two challenges. 
\textbf{Challenge 1}: How to incorporate collaborative information. 
Without explicit labels indicating which users are similar, which users' information should be selected to help personalize generation for the current user?
\textbf{Challenge 2}: 
How to retrieve documents that support personalized LLM generation, rather than relying on traditional semantic relevance?
Pre-trained dense retrieval models~\cite{zhao2024dense} only retrieve based on the semantic relevance between the query and document. Directly using these models for retrieval 
may not necessarily result in content that allows the LLM to generate outputs that meet the user's needs~\cite{shi2024replug,linra}.

To address the above challenges, this paper proposes a method named \textbf{\ourname} which 
adapts \textbf{C}ollaborative \textbf{F}iltering to personalized \textbf{R}etrieval \textbf{A}ugmented \textbf{G}eneration.
Firstly, to address Challenge 1, since there are no explicit user similarity labels, we use contrastive learning~\cite{jaiswal2020survey,wu2020clear} to train user embeddings for retrieving similar users to introduce collaborative information.
Specifically, we apply different data augmentation methods to the user's history to obtain different views, and then treat different views of the same user's history as positive samples for each other. Then we use contrastive learning on different views to train the user embeddings.
Secondly, for Challenge 2, we designed a personalized retriever and reranker to retrieve the top-$k$ documents from the histories of the retrieved users. 
In both retrieval and reranking, in addition to the semantic relevance between the query and documents, we also considered the user's preferences for different documents to enable personalized retrieval.
Additionally, we further fine-tune the retriever and reranker based on the feedback from the LLM to ensure that the retrieved documents better support the personalized LLM generation.
Finally, the top-$k$ documents are concatenated with the user's input query to form a prompt, which is then fed into the LLM for personalized generation.

The major contributions of the paper are summarized as follows:

\noindent\textbf{$\bullet $}~We analyzed the necessity of introducing collaborative filtering into RAG for LLM personalization and identified the challenges: how to introduce collaborative information and how to retrieve documents that support personalized LLM generation.

\noindent\textbf{$\bullet $}~ We proposed a method called \ourname, which uses contrastive learning to train user embeddings for retrieving similar users and incorporating collaborative information. 
% Then, 
It leverages LLM feedback to train the personalized retriever and reranker, enabling them to retrieve documents that support personalized LLM generation.

\noindent\textbf{$\bullet $}~Experimental results on the Language Model Personalization (LaMP)~\cite{salemi2023lamp} benchmark validate the effectiveness of \ourname. The experimental analysis also demonstrates the importance of leveraging collaborative information.

\begin{figure*}[t]
    \centering
        \includegraphics[width=0.93\textwidth]{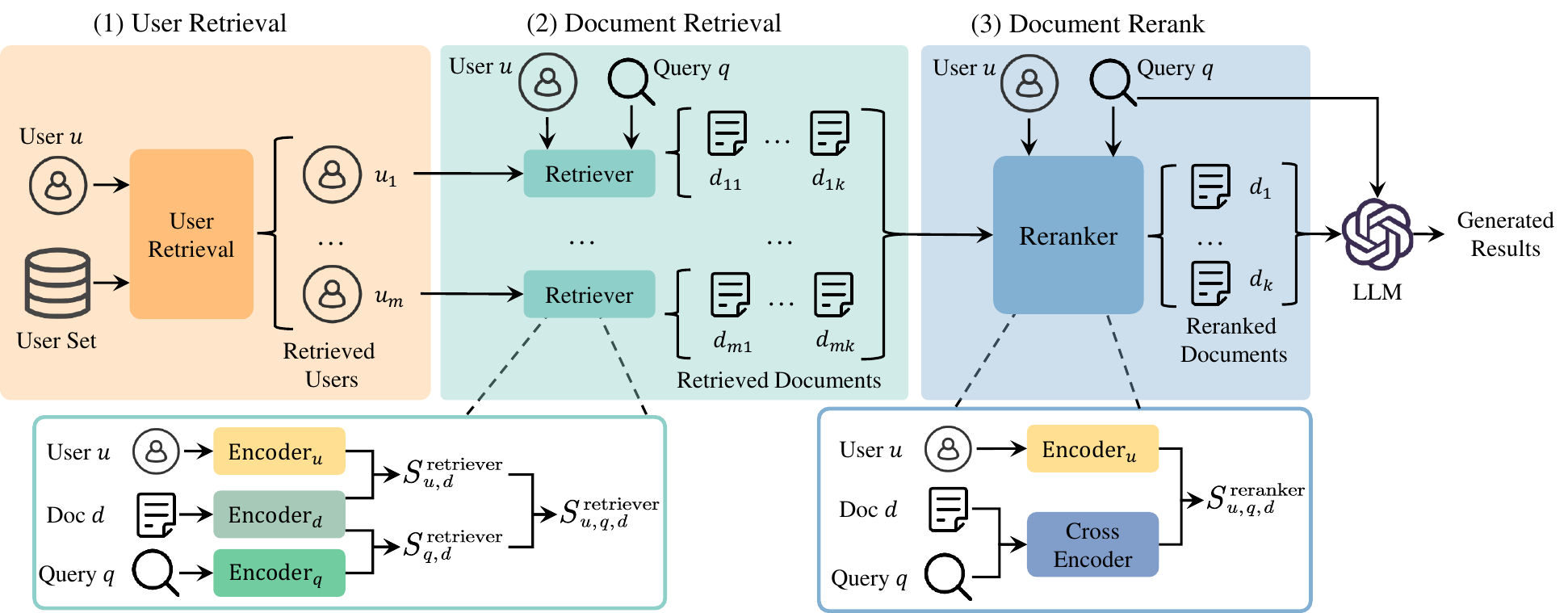}
   \vspace{-10px}
    \caption{
    The architecture of \ourname. From left to right: (a)~User Retrieval retrieves similar users (Section~\ref{sec:user_retrieval});  (b) Retriever retrieves the top-$k$ documents from each user's history (Section~\ref{sec:doc_retrieval}); (c) Reranker reranks the $m\times k$ documents to get the final top-$k$ documents, which are then concatenated with the query and input into the LLM for personalized text generation (Section~\ref{sec:doc_rerank}).
    }
\label{fig:method-model}
\vspace{-0.3cm}
\end{figure*}

\section{Related Work}

\paratitle{Personalization of LLMs.}
Large Language Models (LLMs)~\cite{zhao2023survey} have demonstrated remarkable capabilities in various fields, such as text generation~\cite{li2024pre}, information retrieval~\cite{zhu2023large}, recommender systems~\cite{wu2024survey,dai2023uncovering}, and so on. However, since LLMs are typically designed to serve all tasks with a single model and are trained on broad, domain-agnostic data, they face challenges in adapting to the personalized needs of individual users~\cite{salemi2023lamp,chen2024large}.
Therefore, LLM personalization has attracted widespread attention~\cite{salemi2024optimization,zhuang2024hydra,jang2023personalized}.

Existing works on LLM personalization mainly include the following types of methods: 
(1)~Fine-tuning a personalized LLM for each user~\cite{tan2024democratizinglargelanguagemodels,wu2024fedlora,tan2024personalized}; 
\citet{tan2024democratizinglargelanguagemodels} fine-tuned the LLM using LoRA~\cite{hulora} to get personalized LoRA parameters for each user. 
(2)~Aligning LLMs with user-specific preferences through Reinforcement Learning from Human Feedback (RLHF)~\cite{li2024personalized,wu2024fine,jang2023personalized}; 
\citet{jang2023personalized} first trained different parameters for various objectives using RLHF, then merged these parameters based on users' personalized needs.
(3)~Incorporating user-specific context into the prompt~\cite{richardson2023integrating,salemi2023lamp,salemi2024optimization,zhuang2024hydra,mysore2023pearl,li2023teach}. 
\citet{richardson2023integrating} used instruction-tuned LLMs to summarize user history and then incorporated it into prompts for generation.
\citet{salemi2023lamp,salemi2024optimization} used RAG to retrieve relevant documents from user history based on the input query and incorporated them into the prompt.

This paper further introduces collaborative filtering for personalization based on the RAG framework.
Collaborative filtering has already been applied in fields such as recommender systems~
\cite{shi2024unisar,shen2024survey,zhang2024saqrec,zhang2024qagcf,zhang2024modeling,zhang2024model,zhang2025testtimealignmenttrackinguser,tang2025thinkrecommendunleashinglatent}
and has been proven effective. It assumes that users who have interacted with similar items share similar preferences, and recommending items from similar users to the current user can meet their needs. 
Some works~\cite{xue2017deep,he2017neural} learn the collaborative information between users and items through matrix factorization~\cite{koren2009matrix}, while others~\cite{wang2019neural,he2020lightgcn} further explore higher-order collaborative information between users and items using graph neural networks.
The application of collaborative filtering in LLM personalization remains under-explored.

\paratitle{Retrieval Augmented Generation.}
Retrieval Augmented Generation~\cite{gao2023retrieval,fan2024survey} introduces external knowledge through document retrieval, alleviating issues such as LLM hallucinations~\cite{zhang2023siren}, and enhancing LLMs' capabilities in knowledge-intensive tasks~\cite{kandpal2023large}  such as open-domain question answering~\cite{lewis2020retrieval,izacard2022few}.
Some works~\cite{borgeaud2022improving,izacard2021leveraging} encode retrieved documents using separate encoders, and then fuse the results with the language model using cross-attention.
A more common approach is to directly include the retrieved documents in the prompt of the LLM~\cite{guu2020retrieval,lewis2020retrieval,shi2024replug,linra,asaiself}. 
In recent years, this in-context RAG framework has also been applied to LLM personalization, which is personalized by retrieving documents from the user's history~\cite{salemi2023lamp,salemi2024optimization,zhuang2024hydra}. 
This paper introduces collaborative filtering by retrieving similar users' histories for better personalization.

\section{Problem Formulation}
Let $\mathcal{U}=\{u_1,u_2,\ldots,u_M\}$ denotes the set of all users, where $M$ is the number of users. Each user $u \in \mathcal{U}$ has a chronologically ordered history $\mathcal{H}_u = [d_1,d_2,\ldots,d_{N}]$ which includes all her historical documents, where $N$ is the number of documents in the history.
The personalized text generation dataset is $\mathcal{D}=\{(u,q,y)_i\}^{|\mathcal{D}|}_{i=1}$. For each instance, $q$ is the query input by the user $u$ to the LLM, and $y$ is the target output.
Our goal is first to introduce collaborative information by retrieving the top-$m$ most similar users for  user $u$:
$$
\mathcal{U}_{\mathrm{retrieved}} = \{u_1,u_2,\ldots,u_m\}.
$$
Then, we use a retriever to retrieve the top-$k$ documents from each of the $m$ users' histories, resulting in a total of $m \times k$ documents.
$$
\mathcal{D}_{\mathrm{retrieved}}=\{d_{i,j}| i \in \{1,\ldots,m\}, j \in \{1,\ldots,k\}\}.
$$
Finally, we use a reranker to rerank these $m \times k$ documents and obtain the final top-$k$ documents:
$$
\mathcal{D}_{\mathrm{reranked}}=\{d_i|i \in \{1,\ldots,k\}\}.
$$
These top-$k$ documents will be concatenated with the user's query $q$ as a prompt and input into the LLM, enabling it to generate a response that aligns with the target output $y$.

This paper primarily focuses on how to retrieve $\mathcal{U}_{\mathrm{retrieved}}$ to introduce collaborative information, and how to train the retriever and reranker so that they can effectively retrieve documents that support the personalized LLM generation.

\section{Our Approach}
This section introduces our method \ourname. 
\ourname's overall architecture is shown in Figure~\ref{fig:method-model}.
As mentioned in Section~\ref{sec:intro}, to address Challenge 1, i.e., how to introduce collaborative information, we first train user embeddings using contrastive learning to retrieve the top-$m$ most similar users (see Section~\ref{sec:user_retrieval}). For Challenge 2, which involves retrieving documents that support personalized LLM generation, we fine-tune the personalized retriever and reranker using LLM feedback. The retriever first retrieves the top-$k$ documents from the history of each of the $m$ users, resulting in $m \times k$ documents (see Section~\ref{sec:doc_retrieval}). The reranker then reranks these documents to obtain the final top-$k$ documents as input for the LLM (see Section~\ref{sec:doc_rerank}).

\subsection{User Retrieval}
\label{sec:user_retrieval}
\begin{figure}[t]
    \centering
        \includegraphics[width=0.98\columnwidth]{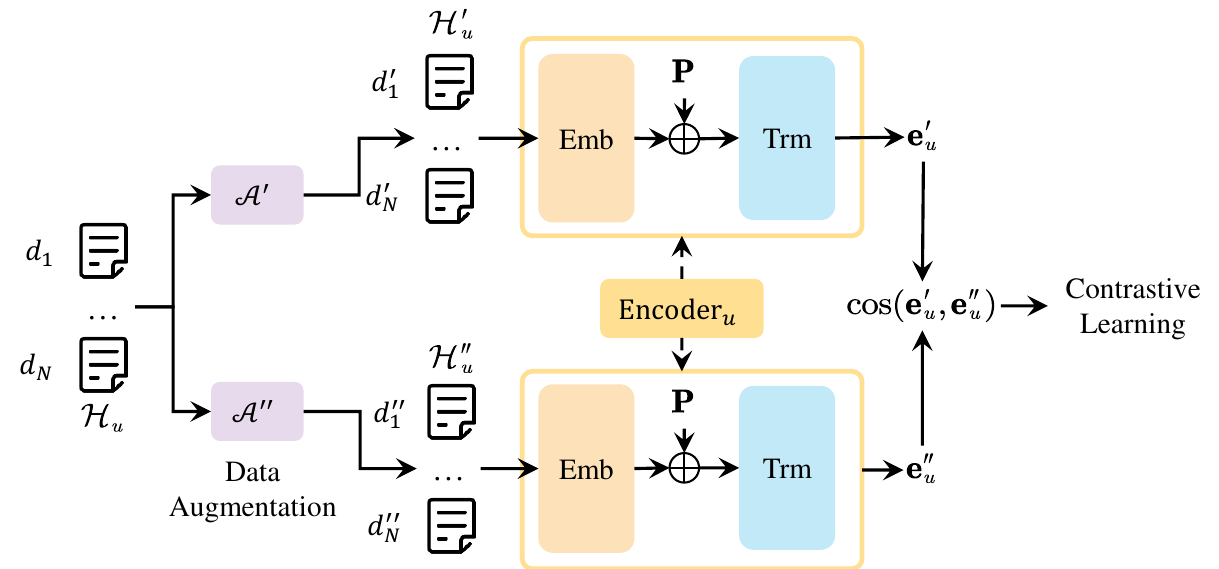}
   \vspace{-8px}
    \caption{Contrastive learning for user embedding training.}
\label{fig:method_user_cl}
\vspace{-0.3cm}
\end{figure}

First, we perform user retrieval to get the top-$m$ most similar users for user $u$ to introduce collaborative information. However, we do not have labels indicating which users are similar to each other. To address this, we employ a contrastive learning~\cite{jaiswal2020survey,wu2020clear} approach. We apply different data augmentation methods to the user history $\mathcal{H}_u$ to obtain different views of the user's history. We treat different views of the same user as positive samples and the histories of other users as negative samples, and then we use the InfoNCE~\cite{oord2018representation} loss to train user embeddings for retrieval.
Figure~\ref{fig:method_user_cl} illustrates the process of training user embeddings using contrastive~learning.

\subsubsection{User Encoder}
\label{sec:user_retrieval:user_encode}
Specifically, we first use an embedding model (such as BERT~\cite{devlin-etal-2019-bert}, RoBERTa~\cite{liu2019roberta}, BGE~\cite{bge_embedding} etc.) $\mathbf{Emb}(\cdot)$ to encode each document in the user's history $\mathcal{H}_u$ to obtain $\mathbf{E}_u=[\mathbf{e}_1,\mathbf{e}_2,\ldots,\mathbf{e}_{N}]^{\intercal} \in \mathbb{R}^{N \times d} $, where $\mathbf{e}_i=\mathbf{Emb}(d_i)$ and $d$ is the embedding dimension.
To model the sequential relationships between different documents in the user's history, 
we introduce positional embedding $\mathbf{P} \in \mathbb{R}^{N \times d}$.
Afterward, the history $\mathcal{H}_u$'s embedding becomes $\widehat{\mathbf{E}}_u=\mathbf{E}_u+\mathbf{P}$.
Then, we apply a transformer~\cite{vaswani2017attention} as the user encoder to encode the user's history $\widehat{\mathbf{E}}_u$ and average the transformer's output to obtain the user's embedding:
\begin{equation}
\label{eq:user_retrieval:user_encode}
    \mathbf{e}_u= \mathrm{Encoder}_u(u)=\mathrm{MEAN}(\mathrm{Trm}(\widehat{\mathbf{E}}_u)) \in \mathbb{R}^{d},
\end{equation}
where $\mathrm{Encoder}_u(\cdot) \rightarrow \mathbb{R}^d$ denotes the user encoder, $\mathrm{Trm}(\cdot)$ denotes a transformer encoder. 
Next, we train the transformer encoder using contrastive learning.

\subsubsection{Data Augmentation}
\label{sec:user_retrieval:data_augment}
We generate different views of $\mathcal{H}_u$ using the following three data augmentation methods:

\textbf{Document Crop.}
We randomly select a continuous sub-sequence of length $L_c=\lfloor \eta_c N \rfloor$ from $\mathcal{H}_u$, where $\eta_c$ is a hyper-parameter controlling the crop ratio. The history after cropping is as follows:
\begin{equation*}
    \mathcal{H}_u^{\mathrm{crop}}=[d_c,d_{c+1},\ldots,d_{c+L_c-1}].
\end{equation*}

\textbf{Document Mask.}
For the history $\mathcal{H}_u$, we randomly mask out $L_m=\lfloor \eta_m N\rfloor$ documents $\mathcal{I}_{\mathrm{mask}}=\{i_1,i_2,\ldots,i_{L_m}\}$, where $\mathcal{I}_{\mathrm{mask}}$ is the set of indices corresponding to the masked documents and $\eta_m$ is a hyper-parameter that controls the mask ratio. 
The masked documents are replaced with a special token [mask].
The history after masking is as follows:
\begin{equation*}
\begin{aligned}
\mathcal{H}_u^{\mathrm{mask}}&=[\hat{d}_1,\hat{d}_2,\ldots,\hat{d}_N], \\ 
\hat{d}_i&=
\begin{cases}
    d_i, & i \notin \mathcal{I}_{\mathrm{mask}}, \\
    [\mathrm{mask}], & i \in \mathcal{I}_{\mathrm{mask}}.  
\end{cases} 
\end{aligned}
\end{equation*}

\textbf{Document Reorder.}
We randomly select a sub-sequence $[d_r,\\d_{r+1},\ldots,d_{r+L_r-1}]$ of length $L_r=\lfloor \eta_r N \rfloor$ from $\mathcal{H}_u$, where $\eta_r$ is a hyper-parameter controlling the reorder ratio, and then randomly shuffle the order of the documents within the sub-sequence to obtain $[\hat{d}_r,\hat{d}_{r+1},\ldots,\hat{d}_{r+L_r-1}]$.
The history after reordering is as follows:
\begin{equation*}
\mathcal{H}_u^{\mathrm{reorder}}=[d_1,d_2,\ldots,\hat{d}_r,\ldots,\hat{d}_{r+L_r-1},\ldots,d_{N}].
\end{equation*}

\subsubsection{Contrastive Loss}
Each time, we randomly select two data augmentation methods $\mathcal{A}^{\prime}$ and $\mathcal{A}^{\prime\prime}$ to generate two different views of $\mathcal{H}_u$, denoted as $\mathcal{H}_u^{\prime}$ and $\mathcal{H}_u^{\prime\prime}$. Then, using the encoder described in Section~\ref{sec:user_retrieval:user_encode}, we obtain the user embeddings $\mathbf{e}_u^{\prime}$ and $\mathbf{e}_u^{\prime\prime}$ corresponding to the different views. Since $\mathbf{e}_u^{\prime}$ and $\mathbf{e}_u^{\prime\prime}$ are obtained through data augmentation of $\mathcal{H}_u$, they are more similar to each other. Therefore, we treat them as positive samples for each other and use the views generated from the augmented histories of other users in the same batch as negative samples. We then perform contrastive learning using the InfoNCE~\cite{oord2018representation} loss as follows: 
\begin{equation}
\label{eq:user_retrieval:cl_loss}
\begin{aligned}
    \mathcal{L}_{\mathrm{CL}}=
    - & \left[ \mathrm{log}\frac{\mathrm{exp}(\mathrm{cos}(\mathbf{e}_u^{\prime},\mathbf{e}_u^{\prime\prime})/\tau_1)}{\sum_{u^{-} \in \mathcal{U}_{\mathrm{neg}}}\mathrm{exp}(\mathrm{cos}(\mathbf{e}_u^{\prime},\mathbf{e}_{u^{-}}^{\prime\prime})/\tau_1)} \right.\\
    & \left. +~~\mathrm{log}\frac{\mathrm{exp}(\mathrm{cos}(\mathbf{e}_u^{\prime},\mathbf{e}_u^{\prime\prime})/\tau_1)}{\sum_{u^{-} \in \mathcal{U}_\mathrm{neg}}\mathrm{exp}(\mathrm{cos}(\mathbf{e}_{u^{-}}^{\prime},\mathbf{e}_{u}^{\prime\prime})/\tau_1)} \right],
\end{aligned}
\end{equation}
where $\tau_1$ is the temperature coefficient, $\mathcal{U}_\mathrm{neg}$ are the set of randomly sampled in-batch negative samples, and $\mathrm{cos}(\cdot)$ denotes the cosine~similarity.

\subsubsection{Top-$m$ User Retrieval}
After training with contrastive learning, we can use the encoder from Section~\ref{sec:user_retrieval:user_encode} to obtain the user embedding $\mathbf{e}_u$. We then calculate the cosine similarity between each pair of user embeddings and retrieve the top-$m$ most similar users $\mathcal{U}_{\mathrm{retrieved}}=\{u_1,u_2,\ldots,u_m\}$ for user $u$. Subsequently, the histories of these $m$ users will be used for further document retrieval.

\subsection{Document Retrieval}
\label{sec:doc_retrieval}
After retrieving the top-$m$ users, we design a personalized retriever to retrieve the top-$k$ documents from each user's history, resulting in a total of $m \times k$ candidate documents $\mathcal{D}_{\mathrm{retrieved}}=\{d_{i,j}| i \in \{1,\ldots,m\}, j \in \{1,\ldots,k\}\}$. 
This section introduces how the retriever is designed and how it's trained to retrieve documents that better align with the requirements of personalized LLM generation.

\subsubsection{Retriever}
First, we use a pre-trained dense retrieval model (such as BGE retriever~\cite{bge_embedding}) to compute the semantic relevance between the query and the candidate documents: 
\begin{equation}
\label{eq:retriever:semantic_score}
    S_{q,d}^{\mathrm{retriever}} = \mathrm{cos}(\mathrm{Encoder}_q(q), \mathrm{Encoder}_d(d)),
\end{equation}
where $\mathrm{Encoder}_q(\cdot) \rightarrow \mathbb{R}^{d}$ and $\mathrm{Encoder}_d(\cdot) \rightarrow \mathbb{R}^{d}$ are the encoders for the query and the document in the retrieval model, respectively. 
Pre-trained retrieval models typically use $S_{q,d}^{\mathrm{retriever}}$ directly for retrieval.
However, $S_{q,d}^{\mathrm{retriever}}$ only considers the semantic relevance between the query and the document. Since different users might input the same query but expect different outputs due to their varying preferences, we further account for user personalization by calculating the preference score of the user for the document as follows:
\begin{equation}
\label{eq:retriever:persona_score}
    S_{u,d}^{\mathrm{retriever}} = \mathrm{cos}(\mathrm{MLP}_1(\mathbf{e}_u), \mathrm{Encoder}_d(d)),
\end{equation}
where $\mathrm{MLP}_1: \mathbb{R}^{d} \rightarrow \mathbb{R}^{d}$ is a multi-layer perceptron that maps the user embedding to the space where the cosine similarity is computed. $\mathbf{e}_u$ is the embedding obtained in Section~\ref{sec:user_retrieval:user_encode}.
The total score for retrieval is computed as follows:
\begin{equation}
\label{eq:retriever:total_score}
    S_{u,q,d}^{\mathrm{retriever}} = (1-\alpha) S_{q,d}^{\mathrm{retriever}} + \alpha S_{u,d}^{\mathrm{retriever}},
\end{equation}
where $\alpha$ is a hyper-parameter that controls the weight of personalization.

\begin{figure}[t]
    \centering
        \includegraphics[width=0.98\columnwidth]{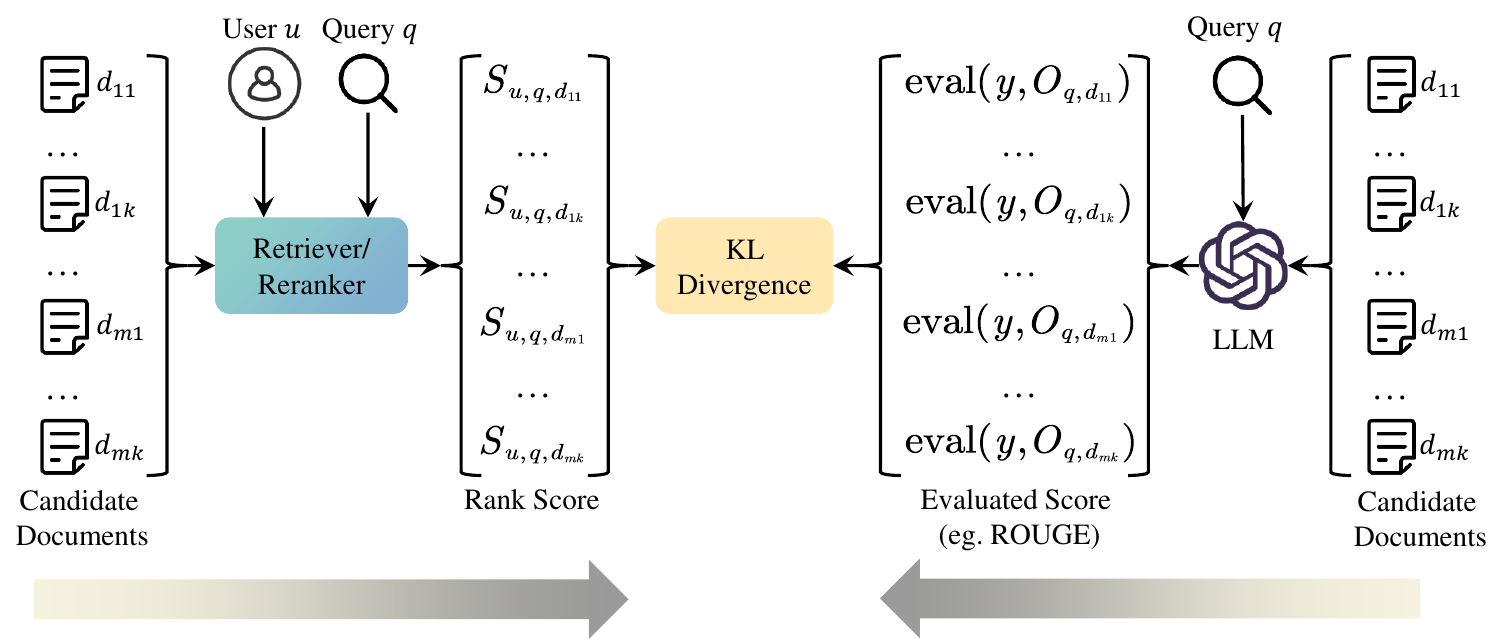}
   \vspace{-10px}
    \caption{The method of training the retriever and reranker using LLM feedback.}
\label{fig:method_loss}
\vspace{-0.3cm}
\end{figure}

\subsubsection{Training}
\label{sec:retriever:training}
Since the pre-trained dense retrieval model is not fine-tuned for our specific task, the retrieved results may not necessarily lead to LLM responses that better match the target output $y$~\cite{shi2024replug,linra}. However, there is no ground truth indicating which documents are better. Therefore, we evaluate the difference between the LLM's output and the target output $y$, using this as a label to train the retrieval model.
Figure~\ref{fig:method_loss} shows the process of training the retriever using LLM feedback.

Specifically, we first use the pre-trained retrieval model to retrieve the top-$k$ documents from each of the $m$ users' histories based on $S_{q,d}^{\mathrm{retriever}}$ in Eq.~\eqref{eq:retriever:semantic_score}, resulting in a total of $m\times k$ candidate documents. These documents are then concatenated with the query one by one and used as prompts for the LLM, producing $m \times k$ outputs:
$$
\{O_{q,d_{i,j}}=\mathrm{LLM}(q,d_{i,j}) | i \in \{1,\ldots,m\}, j \in \{1,\ldots,k\} \},
$$ 
where $\mathrm{LLM}(q,d_{i,j})$ represents the output generated by inputting the concatenated query $q$ and document $d_{i,j}$ into the LLM.
Then, based on the quality of these outputs, we can calculate the distribution of these candidate documents as follows:
\begin{equation}
\label{eq:llm_distribution}
    p_{\mathrm{LLM}}(d_{i,j}|q,y) = \frac{\mathrm{exp}(\mathrm{eval}(y, O_{q,d_{i,j}}))}{\sum_{i=1}^{m}\sum_{j=1}^{k} \mathrm{exp}(\mathrm{eval}(y, O_{q,d_{i,j}}))},
\end{equation}
where $\mathrm{eval}(\cdot)$ measures the difference between the target output $y$ and the LLM's output, using metrics such as ROUGE~\cite{lin2004rouge} score. A larger value returned by $\mathrm{eval}(\cdot)$ indicates a better-generated result.
Similarly, we can also calculate the score distribution of the candidate documents by the retrieval model based on $S_{u,q,d}^{\mathrm{retriever}}$ in Eq.~\eqref{eq:retriever:total_score}:
\begin{equation}
\label{eq:retriever_distribution}
    p_{\mathrm{retriever}}(d_{i,j}|q,u) = \frac{\mathrm{exp}(S_{u,q,d_{i,j}}^{\mathrm{retriever}})}{\sum_{i=1}^{m}\sum_{j=1}^{k} \mathrm{exp}(S_{u,q,d_{i,j}}^{\mathrm{retriever}})}.
\end{equation}
We aim for the retrieval model to retrieve documents that lead to better LLM-generated results, which means making the distribution $p_{\mathrm{retriever}}(d|q,u)$ in Eq.~\eqref{eq:retriever_distribution} closer to the distribution $p_{\mathrm{LLM}}(d|q,y)$ in Eq~\eqref{eq:llm_distribution}. Therefore, we compute the KL divergence between the two distributions as the loss to optimize the retriever:
\begin{equation}
    \mathcal{L}_{\mathrm{retriever}} = \mathrm{KL}(p_{\mathrm{retriever}}(d|q,u)~~||~~p_{\mathrm{LLM}}(d|q,y)).
\end{equation}

\subsection{Document Rerank}
\label{sec:doc_rerank}
After retrieving $\mathcal{D}_{\mathrm{retrieved}}$ through the retriever, in this section, we further refine the results by reranking $\mathcal{D}_{\mathrm{retrieved}}$ to obtain the final top-$k$ ranked results $\mathcal{D}_{\mathrm{reranked}}=\{d_i|i \in \{1,\ldots,k\}\}$.

\subsubsection{Reranker}
\label{sec:reranker}
We use a pre-trained cross-encoder (such as the BGE reranker~\cite{bge_embedding}) to encode the query and document, obtaining the hidden state corresponding to the [CLS] token from the last layer:
\begin{equation}
    \mathbf{h}_{q,d}=\mathrm{CrossEncoder}(q,d),
\end{equation}
where $\mathbf{h}_{q,d} \in \mathbb{R}^{d}$. 
Similarly, when reranking, in addition to considering the semantic relevance between query and document, we also take into account the user's personalized preferences.
However, since the cross-encoder does not encode documents separately, it cannot compute the cosine similarity between users and documents as shown in Eq.~\eqref{eq:retriever:persona_score} to express the user preference score. Therefore, we directly concatenate the user embeddings to the output of the cross-encoder to account for the influence of user preferences.
The overall score used for reranking is calculated as follows:
\begin{equation}
\label{eq:reranker:total_score}
    S_{u,q,d}^{\mathrm{reranker}} = \mathrm{MLP}_3(\mathrm{CONCAT}(\mathbf{h}_{q,d},\mathrm{MLP}_2(\mathbf{e}_u))),
\end{equation}
where $\mathrm{MLP}_2: \mathbb{R}^{d} \rightarrow \mathbb{R}^{d}$ and $\mathrm{MLP}_3: \mathbb{R}^{2d} \rightarrow \mathbb{R}$ are two multi-layer perceptions. $\mathrm{CONCAT}(\cdot)$ denotes the concatenation operation.

\subsubsection{Training}
Similar to the retriever's training in Section~\ref{sec:retriever:training}, we also want the reranker to assign higher scores to the documents that lead to better LLM-generated results. Therefore, we train the reranker using a similar approach. 

We use the trained retrieval model from Section~\ref{sec:retriever:training} to retrieve top-$k$ documents from the history of each of the $m$ users, resulting in a total of $m\times k$ candidate documents. These documents are concatenated with the query $q$ and used as prompts for the LLM, producing $m \times k$ outputs. Similar to Eq.\eqref{eq:llm_distribution}, we can obtain the distribution $p_{\mathrm{LLM}}(d|q,y)$ of these candidate documents.
Based on $S_{u,q,d}^{\mathrm{reranker}}$ in Eq.~\eqref{eq:reranker:total_score}, we can also get the score distribution of the candidate documents by the reranker:
\begin{equation}
\label{eq:reranker_distribution}
    p_{\mathrm{reranker}}(d_{i,j}|q,u) = \frac{\mathrm{exp}(S_{u,q,d_{i,j}}^{\mathrm{reranker}})}{\sum_{i=1}^{m}\sum_{j=1}^{k} \mathrm{exp}(S_{u,q,d_{i,j}}^{\mathrm{reranker}})}.
\end{equation}
We compute the KL divergence between distributions $p_{\mathrm{reranker}}(d|q,u)$ and $p_{\mathrm{LLM}}(d|q,y)$ as the loss to optimize the reranker:
\begin{equation}
    \mathcal{L}_{\mathrm{reranker}} = \mathrm{KL}(p_{\mathrm{reranker}}(d|q,u)~~||~~p_{\mathrm{LLM}}(d|q,y)).
\end{equation}
The loss allows the reranker to assign higher scores to documents that enable better personalized generation by the LLM.

\subsection{Discussion}
\textbf{Computational Efficiency.}
\ourname comprises three modules. The User Encoder is a lightweight, single-layer Transformer with inputs derived from a frozen BGE embedding (dimension 768), resulting in minimal parameter overhead. The retriever and reranker are comparable in size to BERT (approximately 100M parameters). Overall, the training cost is low due to the modest parameter size. During inference, user and document embeddings can be precomputed, requiring only similarity calculations for retrieval, ensuring minimal computational cost. This efficiency enables our method to generalize quickly to new datasets.

\begin{table}[t]
    \caption{Statistics of the datasets used in this paper.}
    \vspace{-8px}
    \center
     \resizebox{.98\columnwidth}{!}{
        \begin{tabular}{ccccccc}
        \toprule
        Dataset & LaMP-1 & LaMP-2 & LaMP-3 & LaMP-4 & LaMP-5 & LaMP-7  \\
        \midrule
        \#Users & 6,542 & 929 & 20,000 & 1,643 & 14,682 & 13,437 \\
        \#Train & 6,542 & 5,073 & 20,000 & 12,500 & 14,682 & 13,437 \\
        \#Dev & 1,500 & 1,410 & 2,500 & 1,500 & 1,500 & 1,498 \\
        \#Test & 1,500 & 1,557 & 2,500 & 1,800 & 1,500 & 1,500 \\
        \bottomrule
        \end{tabular}}
    \label{tab:dataStatistics}   
   \vspace{-0.3cm}
\end{table}

\begin{table*}[t]
\centering
\caption{
Comparison of the performance of \ourname with other approaches on the LaMP benchmark.
$\uparrow$ indicates that a higher value for the corresponding metric is better, while $\downarrow$ indicates that a lower value is better.
The best and the second-best methods are highlighted in bold and underlined fonts, respectively.
``*'' indicates improvements over the second-best methods are statistically significant ($t$-test, $p$-value$<0.05$).
}
\vspace{-3px}
\label{tab:main_result}
\resizebox{.98\textwidth}{!}{
\begin{tabular}{llcccccccccccc}
\toprule
\multicolumn{1}{l}{\multirow{2}{*}{LLMs}} & 
\multicolumn{1}{l}{\multirow{2}{*}{Retrievers}} & 
\multicolumn{2}{c}{LaMP-1} & 
\multicolumn{2}{c}{LaMP-2} & 
\multicolumn{2}{c}{LaMP-3} & 
\multicolumn{2}{c}{LaMP-4} & 
\multicolumn{2}{c}{LaMP-5} & 
\multicolumn{2}{c}{LaMP-7} \\
\cmidrule(l){3-4} \cmidrule(l){5-6} \cmidrule(l){7-8} 
\cmidrule(l){9-10} \cmidrule(l){11-12} \cmidrule(l){13-14}
\multicolumn{1}{c}{}  & \multicolumn{1}{c}{} 
&Accuracy~$\uparrow$ &F1~$\uparrow$ &Accuracy~$\uparrow$ &F1~$\uparrow$ &MAE~$\downarrow$ &RMSE~$\downarrow$ &ROUGE-1~$\uparrow$ &ROUGE-L~$\uparrow$ &ROUGE-1~$\uparrow$ &ROUGE-L~$\uparrow$ &ROUGE-1~$\uparrow$ &ROUGE-L~$\uparrow$ \\
\midrule
\multirow{7}*{Llama3}
&Zero Shot &0.4993 &0.2497 &0.2993 &0.0200 &0.5024 &0.7904 &0.1406 &0.1228 &0.4417 &0.3650 &0.3079 &0.2593 \\
&Random &0.5740 &0.2870 &0.3929 &0.0262 &0.4104 &0.7833 &0.1787 &0.1571 &0.4533 &0.3875 &0.3137 &0.2508 \\
&Recency &0.6040 &0.3020 &0.3993 &0.0266 &0.3980 &0.7491 &\underline{0.1856} &\underline{0.1650} &0.4573 &0.3928 &0.3325 &0.2686 \\
&BM25~\cite{robertson1995okapi} &0.6240 &0.3120 &0.4255 &0.0284 &0.4060 &0.7666 &0.1803 &0.1591 &0.4637 &\underline{0.3978} &0.3449 &0.2780 \\
&BGE~\cite{bge_embedding} 
&0.6327 &0.3163 &0.4574 &0.0305 &0.3528 &0.6969 &0.1811 &0.1611 &0.4638 &0.3958 &0.3391 &0.2742 \\
&ROPG~\cite{salemi2024optimization} &\underline{0.6440} &\underline{0.3220} &\underline{0.4681} &\underline{0.0312} &\underline{0.3456} &\underline{0.6922} &0.1838 &0.1634 &\underline{0.4638} &0.3956 &\underline{0.3530} &\underline{0.2881} \\
&\textbf{\ourname} &\textbf{0.6533}* &\textbf{0.3267}* &\textbf{0.5340}* &\textbf{0.0356}* &\textbf{0.2812}* &\textbf{0.5997}* &\textbf{0.1957}* &\textbf{0.1745}* &\textbf{0.4810}* &\textbf{0.4153}* &\textbf{0.3752}* &\textbf{0.3055}* \\
\hline
\multirow{7}*{Qwen2}
&Zero Shot &0.5000 &0.2500 &0.2908 &0.0194 &0.4444 &0.7805 &0.1264 &0.1081 &0.4144 &0.3468 &0.3972 &0.3229 \\
&Random &0.5633 &0.2817 &0.3284 &0.0219 &0.4000 &0.7621 &0.1581 &0.1377 &0.4580 &0.3921 &0.4291 &0.3564 \\
&Recency &0.5773 &0.2887 &0.3326 &0.0222 &0.3912 &0.7563 &0.1581 &0.1369 &0.4562 &0.3913 &0.4247 &0.3525 \\
&BM25~\cite{robertson1995okapi} &0.5987 &0.2993 &0.3532 &0.0235 &0.4228 &0.8027 &0.1580 &0.1374 &\underline{0.4613} &\underline{0.3950} &0.4290 &0.3570 \\
&BGE~\cite{bge_embedding} &0.6080 &0.3040 &0.3674 &0.0245 &0.3696 &\underline{0.7211} &0.1613 &0.1398 &0.4571 &0.3910 &\underline{0.4347} &0.3605 \\

&ROPG~\cite{salemi2024optimization} &\underline{0.6093} &\underline{0.3047} &\underline{0.3830} &\underline{0.0255} &\underline{0.3672} &0.7332 &\underline{0.1617} &\underline{0.1401} &0.4600 &0.3946 &0.4345 &\underline{0.3610} \\

&\textbf{\ourname} &\textbf{0.6133} &\textbf{0.3067} &\textbf{0.3957}* &\textbf{0.0264} &\textbf{0.3536}* &\textbf{0.7071}* &\textbf{0.1621} &\textbf{0.1412} &\textbf{0.4703}* &\textbf{0.4029}* &\textbf{0.4425}* &\textbf{0.3708}* \\
\bottomrule
\end{tabular} 
}
\vspace{-0.3cm}
\end{table*}

\section{Experiments}
We conducted experiments to evaluate the performance of \ourname. 
The source code is available. 
\footnote{\url{https://github.com/TengShi-RUC/CFRAG}}

\subsection{Experimental Setup}

\subsubsection{Dataset} 
We conducted experiments on the Language Model Personalization (LaMP)~\cite{salemi2023lamp} benchmark, which consists of seven personalized text generation tasks. We excluded LaMP-6 because its data is not publicly available. The remaining tasks include:
\textbf{LaMP-1}~(Personalized Citation Identification);
\textbf{LaMP-2}~(Personalized Movie Tagging);
\textbf{LaMP-3}~(Personalized Product Rating);
\textbf{LaMP-4}~(Personalized News Headline Generation);
\textbf{LaMP-5}~(Personalized Scholarly Title Generation);
\textbf{LaMP-7}~(Personalized Tweet Paraphrasing).
We used the time-based split provided by LaMP to divide the data into training, validation, and test sets.
The statistics of these datasets are shown in Table~\ref{tab:dataStatistics}.

\subsubsection{Evaluation Metrics}
Following previous works~\cite{salemi2023lamp,salemi2024optimization}, we evaluate Accuracy and F-1 score for LaMP-1 and LaMP-2, mean absolute error (MAE) and root mean squared error (RMSE) for LaMP-3, ROUGE-1 and ROUGE-L~\cite{lin2004rouge} for LaMP-4, LaMP-5 and~LaMP-7.

\subsubsection{Baselines}
In this work, we compare \ourname with the following methods.

\textbf{No Personalization}: 
We directly input the user's query into the LLM without retrieving from user history, using this as the non-personalized baseline. We refer to this method as \textbf{Zero Shot}.

\textbf{Personalized Baselines}: 
We compared \ourname with methods that personalize by retrieving from user history using different retrieval models, including:
(1)~\textbf{Random} selects $k$ items randomly from the user's history;
(2)~\textbf{Recency} selects the most recent $k$ items from the user's history;
(3)~\textbf{BM25}~\cite{robertson1995okapi} retrieves top-$k$ items from the user's history using BM25;
(4)~\textbf{BGE}~\cite{bge_embedding} retrieves top-$k$ items from the user's history using BGE retriever;
(5)~\textbf{ROPG}~\cite{salemi2024optimization} optimizes the dense retrieval model based on the results generated by the LLM.

\subsubsection{Implementation Details}
We conducted experiments on two LLMs: Llama3-8B-Instruct~\cite{llama3modelcard} and Qwen2-7B-Instruct~\cite{yang2024qwen2}.
In this paper, we do not fine-tune the LLM because fine-tuning is costly and could cause the LLM to retain user information, potentially compromising user privacy.
To ensure a fair comparison, we use greedy search for text generation.
The dense retrieval model used in all methods is bge-base-en-v1.5\footnote{\url{https://huggingface.co/BAAI/bge-base-en-v1.5}}~\cite{bge_embedding}.
The cross-encoder used for reranker in Section~\ref{sec:reranker} is bge-reranker-base\footnote{\url{https://huggingface.co/BAAI/bge-reranker-base}}~\cite{bge_embedding}.
All hyper-parameters for the baselines are searched according to the settings in the original papers. 
The embedding dimension $d$ is set to 768.
The number of retrieved documents $k$ is set to 5, and the number of retrieved users $m$ is tuned among $\{2,3,4,5,6\}$.
The $\mathrm{Trm}(\cdot)$ encoder in Eq.~\eqref{eq:user_retrieval:user_encode} has 1 layer and 2 heads.
The hyper-parameters $L_c$, $L_m$, and $L_r$ used for data augmentation in Section~\ref{sec:user_retrieval:data_augment} are set to 0.7, 0.3, and 0.3, respectively.
The temperature parameters $\tau_1$ in Eq.~\eqref{eq:user_retrieval:cl_loss} is tuned among $\{0.01,0.1,1\}$.
The weight $\alpha$ in Eq.~\eqref{eq:retriever:total_score} is tuned among $[0.01,1.0]$.
The learning rate is tuned among $\{1e\text{-}3,1e\text{-}4,1e\text{-}5\}$.
Adam \cite{kingma2014adam} is used to conduct the~optimization.
The data input and output formats are provided in Appendix~\ref{appendix:prompt}.

\begin{table*}[h!]
\centering
\caption{
Ablation Study of \ourname on LaMP based on Llama3. ``MEAN'' represents using the average of user history document embeddings as the user embedding. ``w/o'' indicates the corresponding module in \ourname is removed.
}
\vspace{-5px}
\label{tab:ablation_result}
\resizebox{.98\textwidth}{!}{
\begin{tabular}{clcccccccccccc}
\toprule
\multicolumn{2}{c}{Variants} & 
\multicolumn{2}{c}{LaMP-1} & 
\multicolumn{2}{c}{LaMP-2} & 
\multicolumn{2}{c}{LaMP-3} & 
\multicolumn{2}{c}{LaMP-4} & 
\multicolumn{2}{c}{LaMP-5} & 
\multicolumn{2}{c}{LaMP-7} \\
\cmidrule(l){1-2}
\cmidrule(l){3-4} \cmidrule(l){5-6} \cmidrule(l){7-8} 
\cmidrule(l){9-10} \cmidrule(l){11-12} \cmidrule(l){13-14}
\# &Model 
&Accuracy~$\uparrow$ &F1~$\uparrow$ &Accuracy~$\uparrow$ &F1~$\uparrow$ &MAE~$\downarrow$ &RMSE~$\downarrow$ &ROUGE-1~$\uparrow$ &ROUGE-L~$\uparrow$ &ROUGE-1~$\uparrow$ &ROUGE-L~$\uparrow$ &ROUGE-1~$\uparrow$ &ROUGE-L~$\uparrow$ \\
\midrule

(0) &\textbf{\ourname} &\textbf{0.6533} &\textbf{0.3267} &\textbf{0.5340} &\textbf{0.0356} &\textbf{0.2812} &\textbf{0.5997} &\textbf{0.1957} &\textbf{0.1745} &\textbf{0.4810} &\textbf{0.4153} &\textbf{0.3752} &\textbf{0.3055} \\
\hdashline

(1) &w/o User Retrieval &0.6400 &0.3200 &0.4936 &0.0329 &0.3444 &0.6925 &0.1914 &0.1689 &0.4642 &0.3963 &0.3566 &0.2903 \\
(2) &User Retrieval (MEAN) &0.6420 &0.3210 &0.5064 &0.0338 &0.3412 &0.6867 &0.1847 &0.1639 &0.4779 &0.4113 &0.3722 &0.3022 \\
\hdashline
(3) &w/o Retriever Tuning &0.6453 &0.3227 &0.4979 &0.0332 &0.2852 &0.6070 &0.1916 &0.1704 &0.4742 &0.4048 &0.3599 &0.2940 \\
(4) &w/o $S^{\mathrm{retriever}}_{u,d}$ in Eq.~\eqref{eq:retriever:total_score} &0.6333 &0.3167 &0.5113 &0.0341 &0.3324 &0.6861 &0.1895 &0.1696 &0.4750 &0.4088 &0.3732 &0.3039 \\
\hdashline
(5) &w/o Reranker Tuning &0.6307 &0.3153 &0.4695 &0.0313 &0.3696 &0.7392 &0.1766 &0.1550 &0.4714 &0.4068 &0.3432 &0.2775 \\
(6) &w/o $\mathbf{e}_u$ in Eq.~\eqref{eq:reranker:total_score} &0.6313 &0.3157 &0.4993 &0.0333 &0.3420 &0.6925 &0.1887 &0.1672 &0.4772 &0.4123 &0.3731 &0.3030 \\
\bottomrule
\end{tabular} 
}
\vspace{-0.3cm}
\end{table*}

\subsection{Experimental Results}
Experimental results are shown in Table~\ref{tab:main_result}. From the results, we can find that:

\noindent\textbf{$\bullet $}~Firstly, compared to existing methods, \ourname achieved the best results across six datasets in the LaMP benchmark. This demonstrates the effectiveness of introducing collaborative information between users into RAG and using LLM feedback to tune the retriever and reranker to ensure that they can retrieve the documents that support the personalized LLM generation.

\noindent\textbf{$\bullet $}~Secondly, we can observe that even randomly selecting user history outperforms the zero-shot method without any user history. This highlights the importance of incorporating user history to reflect user preferences for personalized generation. Additionally, we observe that retrieval methods perform better than simply selecting the most recent user history, underscoring the importance of retrieval.

\noindent\textbf{$\bullet $}~Thirdly, we also observe that, in most cases, RAG and ROPG methods using dense retrieval models outperform BM25. Additionally, \ourname, which fine-tunes the retriever based on LLM feedback, achieves better results. This shows, on the one hand, that the better the retriever, the better the generation results, and on the other hand, fine-tuning the retriever based on LLM feedback to ensure it can retrieve the documents that meet the personalized generation needs of LLM is crucial.

\subsection{Ablation Study}
We conducted an ablation study to investigate the effectiveness of different modules in \ourname, as shown in Table~\ref{tab:ablation_result}.
\ourname consists of three modules: User Retrieval, Document Retrieval, and Document Rerank. 
We removed different modules from \ourname one by one to verify the effectiveness of each module.

\subsubsection{User Retrieval}
First, we validated the effectiveness of introducing collaborative information by retrieving similar users, as shown in row (1) of Table~\ref{tab:ablation_result}. It can be seen that without retrieving similar users and only retrieving from the current user's history, the performance is worse than that of \ourname, highlighting the importance of collaborative information.

We also validated the effectiveness of training user embeddings using contrastive learning. For comparison, we directly averaged the document embeddings from the user's history to create user embeddings for retrieval, as shown in row (2) of Table~\ref{tab:ablation_result}. 
It can be seen that \ourname, which uses user embeddings trained with contrastive learning, achieves better results. This is because contrastive learning constructs user similarity labels through data augmentation and uses the InfoNCE loss to help the embeddings learn which users are similar. In contrast, using mean pooling directly cannot capture user similarity.

\begin{figure}[t]
     \centering
     \subfigure[LaMP-1]{
        \label{fig:ret_user_lamp_1}
        \includegraphics[width=0.475\columnwidth]{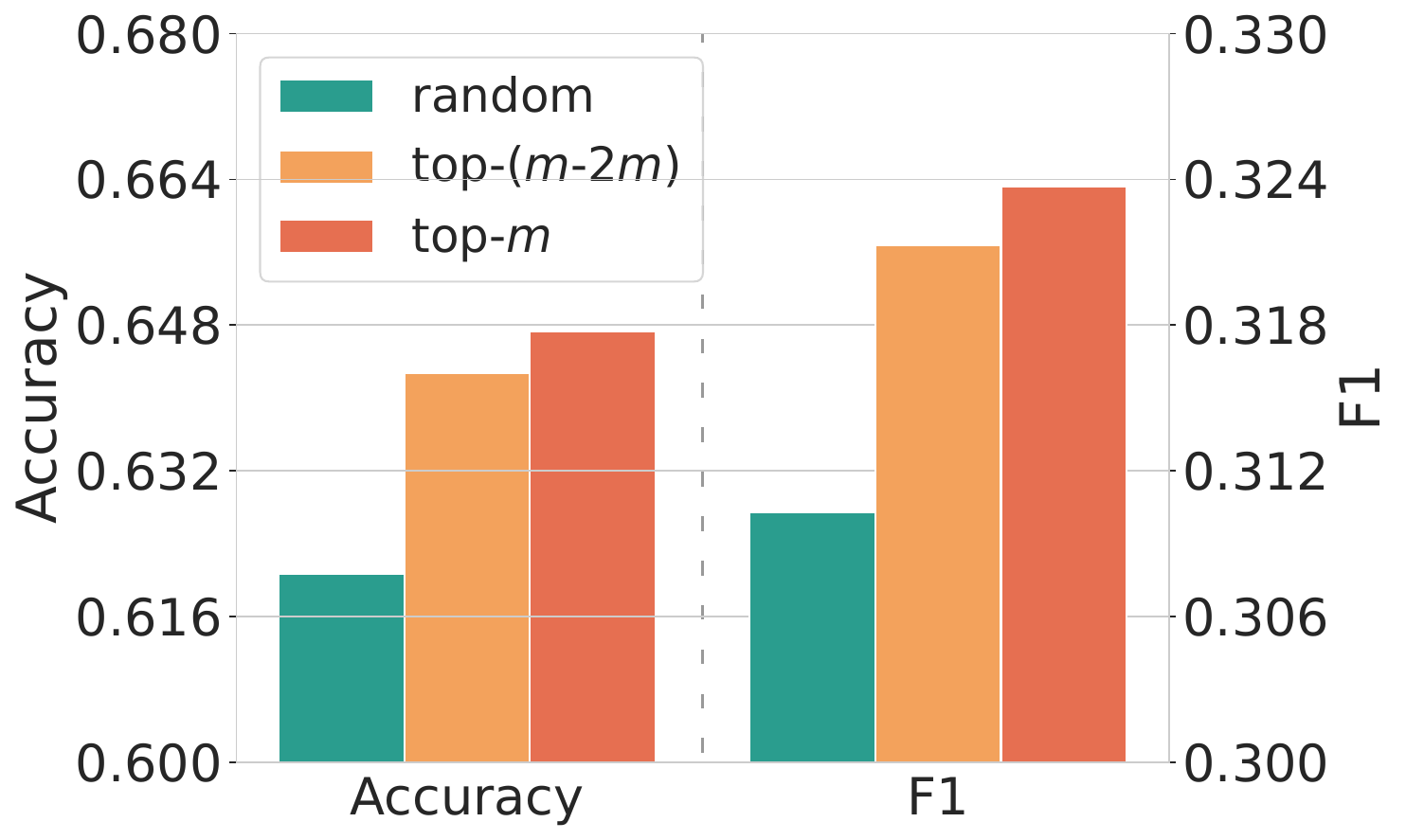}
     }
    \subfigure[LaMP-5]{
        \label{fig:ret_user_lamp_5}
        \includegraphics[width=0.475\columnwidth]{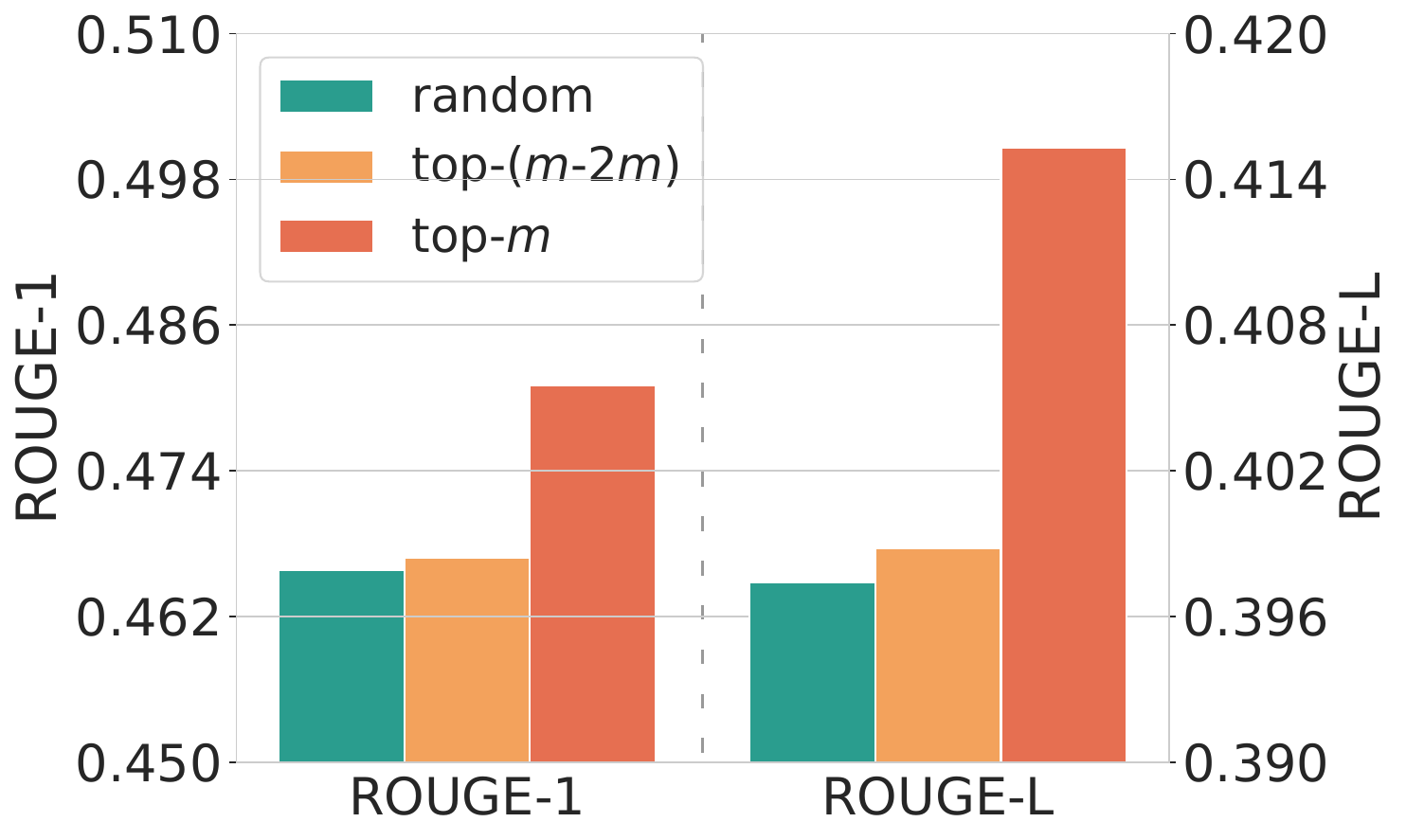}
     }
     \vspace{-5px}
     \caption{
     Results of using different methods to select users for introducing collaborative information. ``random'' indicates randomly selecting $m$ users; ``top-($m$-$2m$)'' represents selecting users whose similarity to the current user ranks between $m$ and $2m$; ``top-$m$'' indicates selecting the most similar $m$ users.
     }
     \label{fig:ret_user}
     \vspace{-0.3cm}
\end{figure}

\subsubsection{Document Retrieval}
We also validated the effectiveness of the personalized retriever we designed, as shown in Table~\ref{tab:ablation_result}, rows (3) and (4). First, in row (3), we can see that without fine-tuning based on LLM feedback, using a pre-trained dense retrieval model leads to worse performance. This indicates that retrieval cannot be based solely on semantic relevance, ensuring that the retrieved documents support personalized LLM generation is crucial.
Additionally, we analyzed the impact of removing $S^{\mathrm{retriever}}_{u,d}$ from Eq.~\eqref{eq:retriever:persona_score} and only using $S^{\mathrm{retriever}}_{q,d}$ from Eq.~\eqref{eq:retriever:semantic_score} for retrieval, as indicated in row (4). The results decreased, demonstrating that users' personalized preferences should also be considered during retrieval, rather than solely focusing on the semantic relevance between the query and documents.

\subsubsection{Document Rerank}
We also validated the effectiveness of the personalized reranker we designed, as shown in Table~\ref{tab:ablation_result}, rows (5) and (6).
First, in row (5), it can be seen that using a pre-trained reranker leads to worse results, highlighting the importance of fine-tuning based on LLM feedback.
We also observed the effect of removing $\mathbf{e}_u$ from Eq.~\eqref{eq:reranker:total_score} and only using $\mathbf{h}_{q,d}$ to calculate $S_{q,d}^{\mathrm{reranker}}$ for ranking, as indicated in row (6). The results decreased in this case, highlighting the importance of considering users' personalized preferences in the reranker.

\subsection{Experimental Analysis}
As mentioned in Section~\ref{sec:intro}, adapting collaborative filtering into personalized RAG faces two challenges.
\textbf{Challenge 1}: How to introduce collaborative information?
\textbf{Challenge 2}: How to retrieve documents that support personalized LLM generation?
In this section, we conduct experimental analysis to further demonstrate the effectiveness of our method in addressing these two challenges. Additionally, we provide further analysis of the results of \ourname and the impact of hyper-parameters. 
Due to space limitations, we conducted experimental analysis on the LaMP-1 and LaMP-5 datasets.

\begin{figure}[t]
     \centering
     \subfigure[LaMP-1]{
        \label{fig:tune_lamp_1}
        \includegraphics[width=0.475\columnwidth]{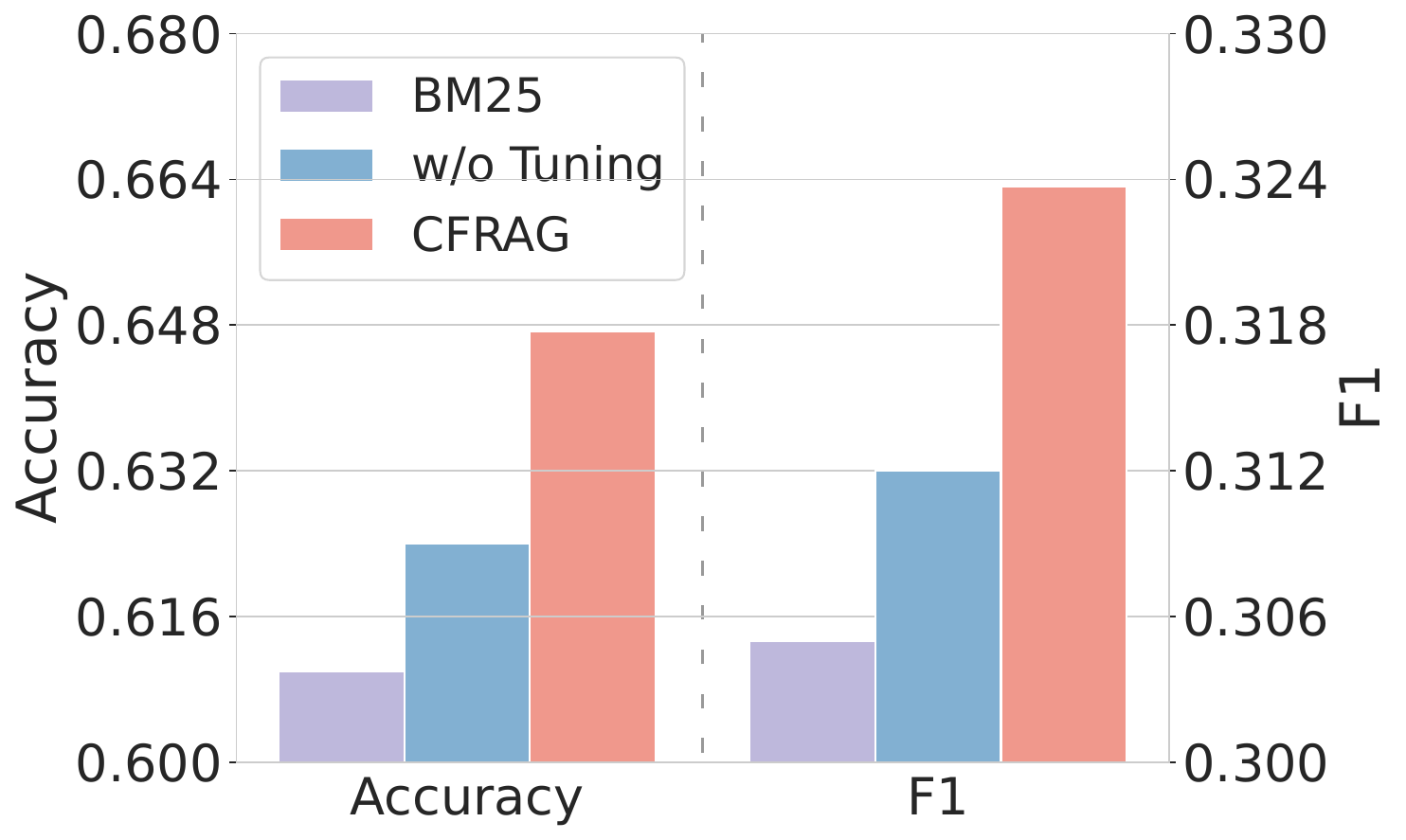}
     }
    \subfigure[LaMP-5]{
        \label{fig:tune_lamp_5}
        \includegraphics[width=0.475\columnwidth]{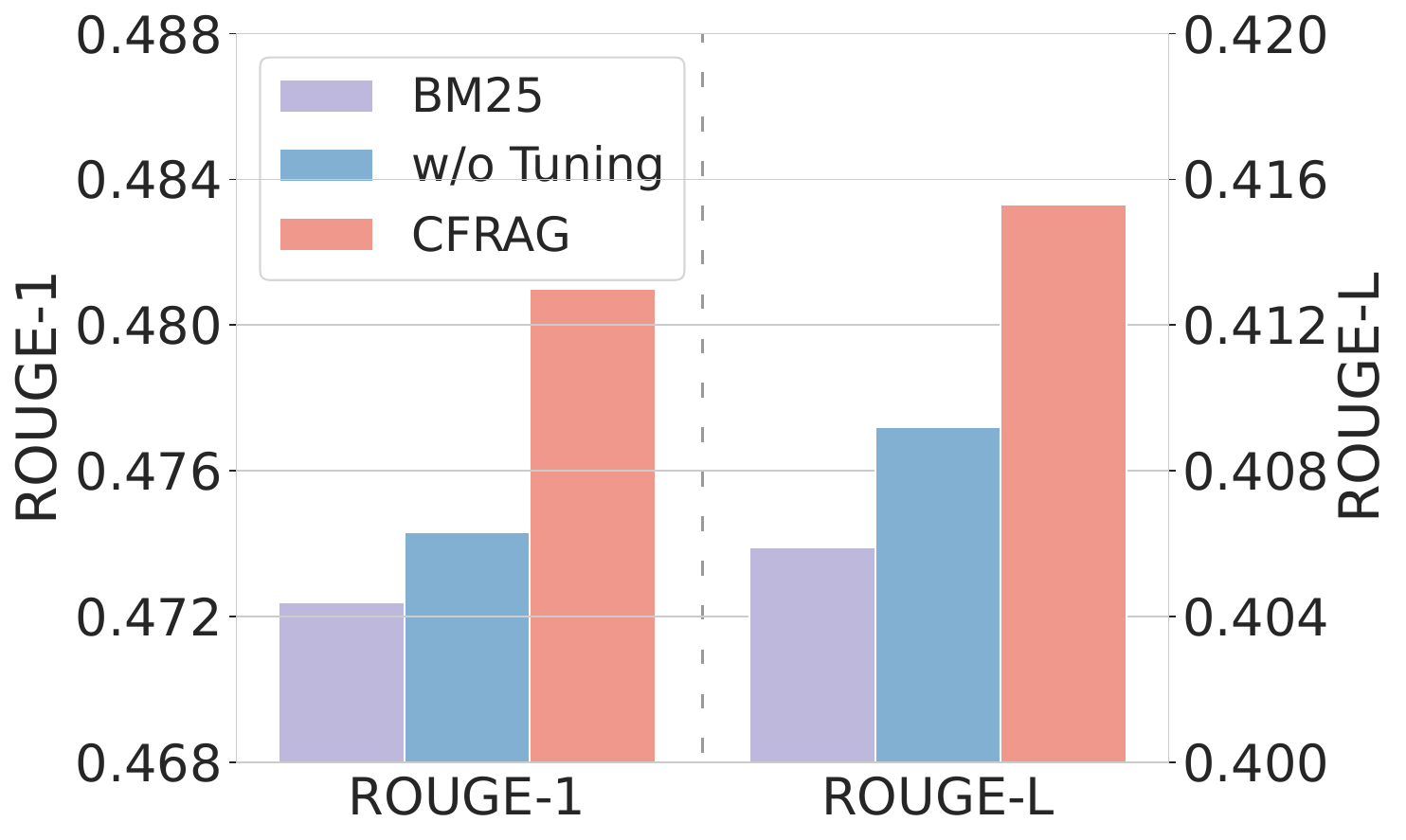}
     }
     \caption{
     Results using different retrievers and rerankers. ``BM25'' indicates using BM25 as both the retriever and reranker, while ``w/o Tuning'' refers to using pre-trained retrievers and rerankers without LLM feedback fine-tuning.
     }
     \label{fig:tune}
     \vspace{-0.3cm}
\end{figure}

\begin{figure}[t]
     \centering
     \subfigure[LaMP-1]{
        \label{fig:doc_ratio_lamp_1}
        \includegraphics[width=0.475\columnwidth]{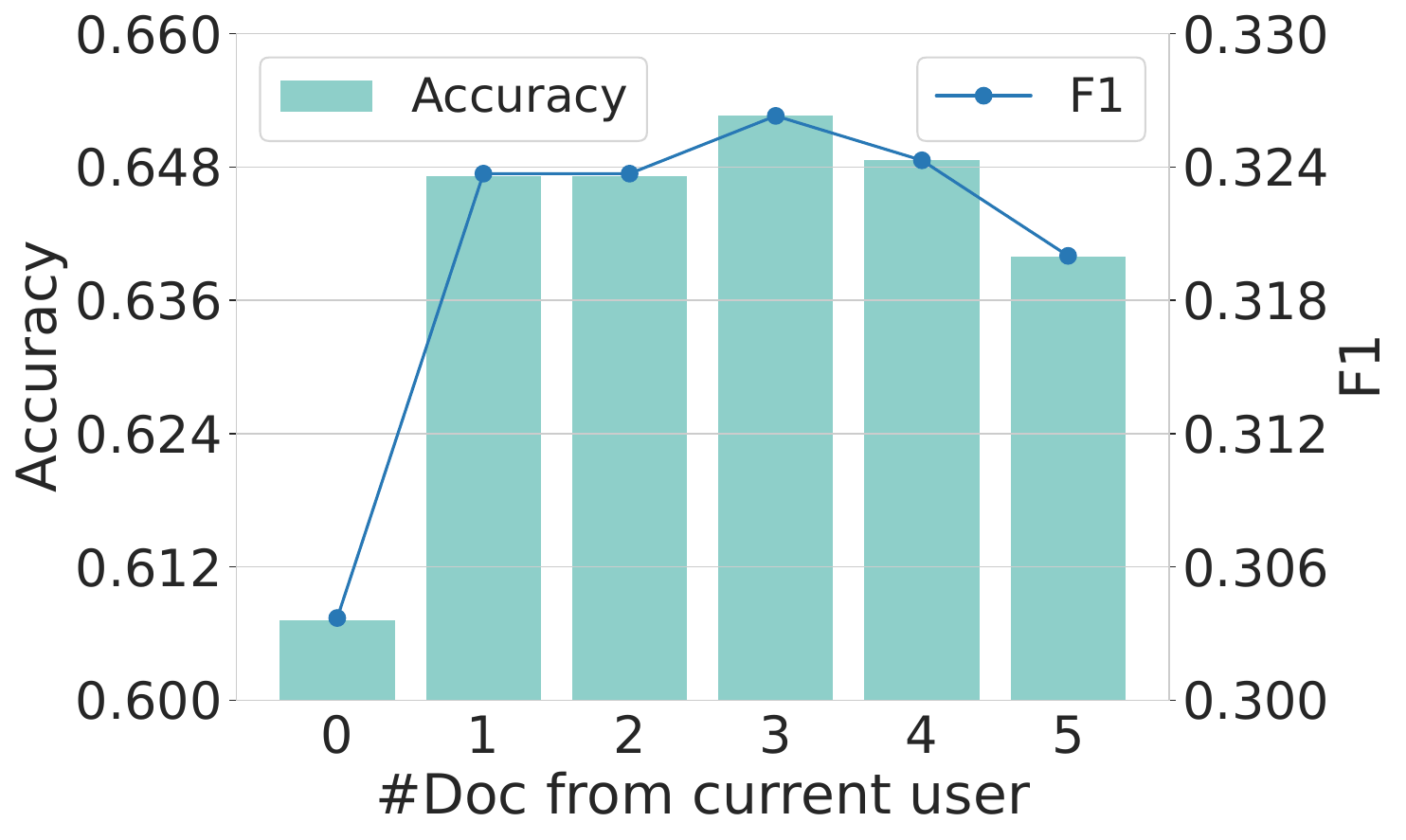}
     }
    \subfigure[LaMP-5]{
        \label{fig:doc_ratio_lamp_5}
        \includegraphics[width=0.475\columnwidth]{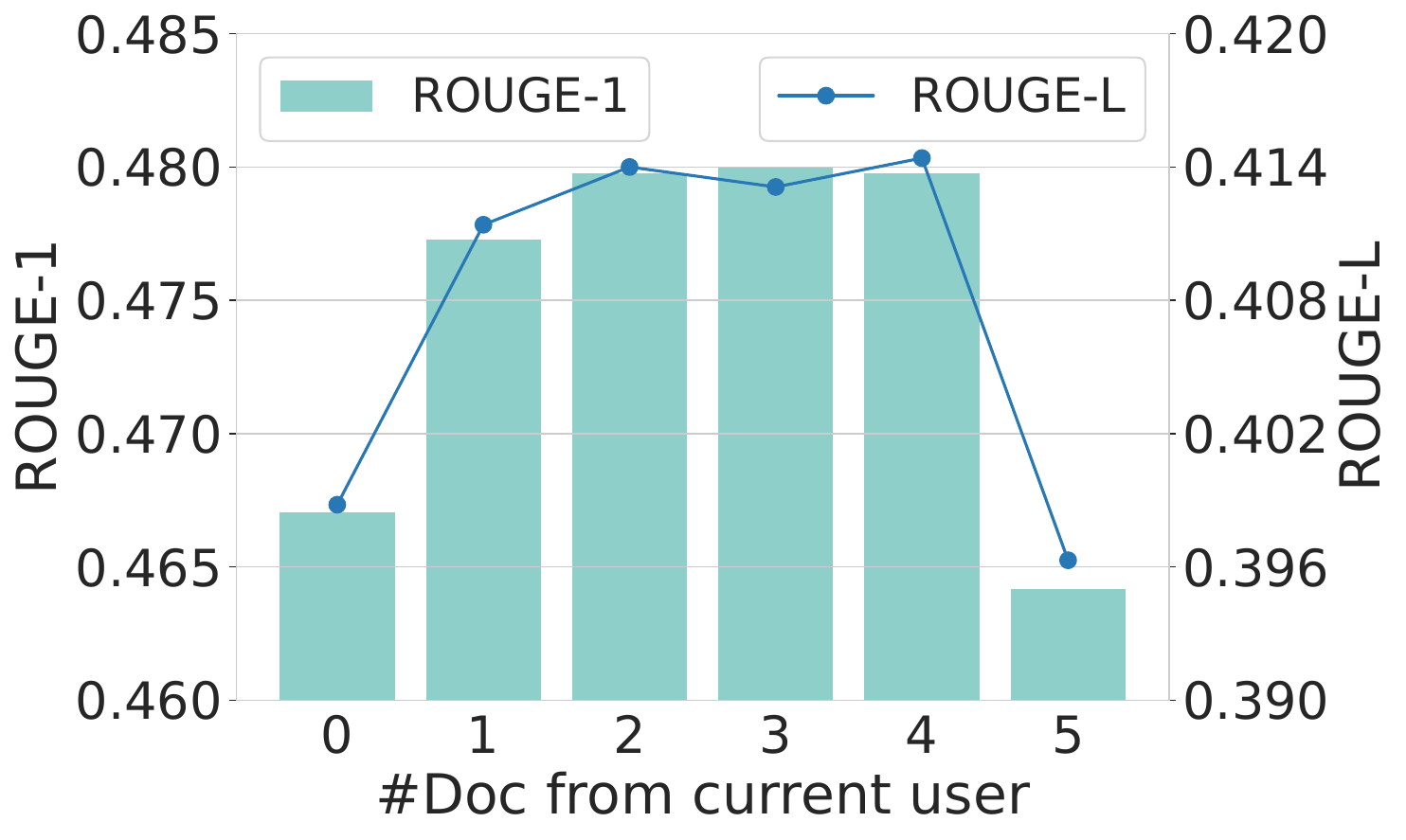}
     }
     \caption{
    Performance under different numbers of retrieved documents from the current user $u$'s history in the top-$k$ documents.
     }
     \label{fig:doc_ratio}
     \vspace{-0.3cm}
\end{figure}

\subsubsection{
Effectiveness of User Retrieval using Contrastive Learning (Challenge 1)
}
As described in Section~\ref{sec:intro}, to address Challenge 1, we train user embeddings using contrastive learning to retrieve the top-$m$ most similar users for introducing collaborative information.
To validate the effectiveness of this approach, we compared it with randomly selecting $m$ users and selecting users from top-$m$ to $2m$, as shown in Figure~\ref{fig:ret_user}. 
First, we can see that randomly selecting users yields the worst performance, indicating that collaborative information cannot be introduced indiscriminately. Secondly, the results show that retrieving users from the range of top-$m$ to $2m$ performs worse than using the top-$m$ users, suggesting that information from users who are more similar to the current user $u$ is more important. These highlight the importance of retrieving the most similar top-$m$ users

\begin{figure}[t]
     \centering
     \subfigure[LaMP-1]{
        \label{fig:n_user_lamp_1}
        \includegraphics[width=0.475\columnwidth]{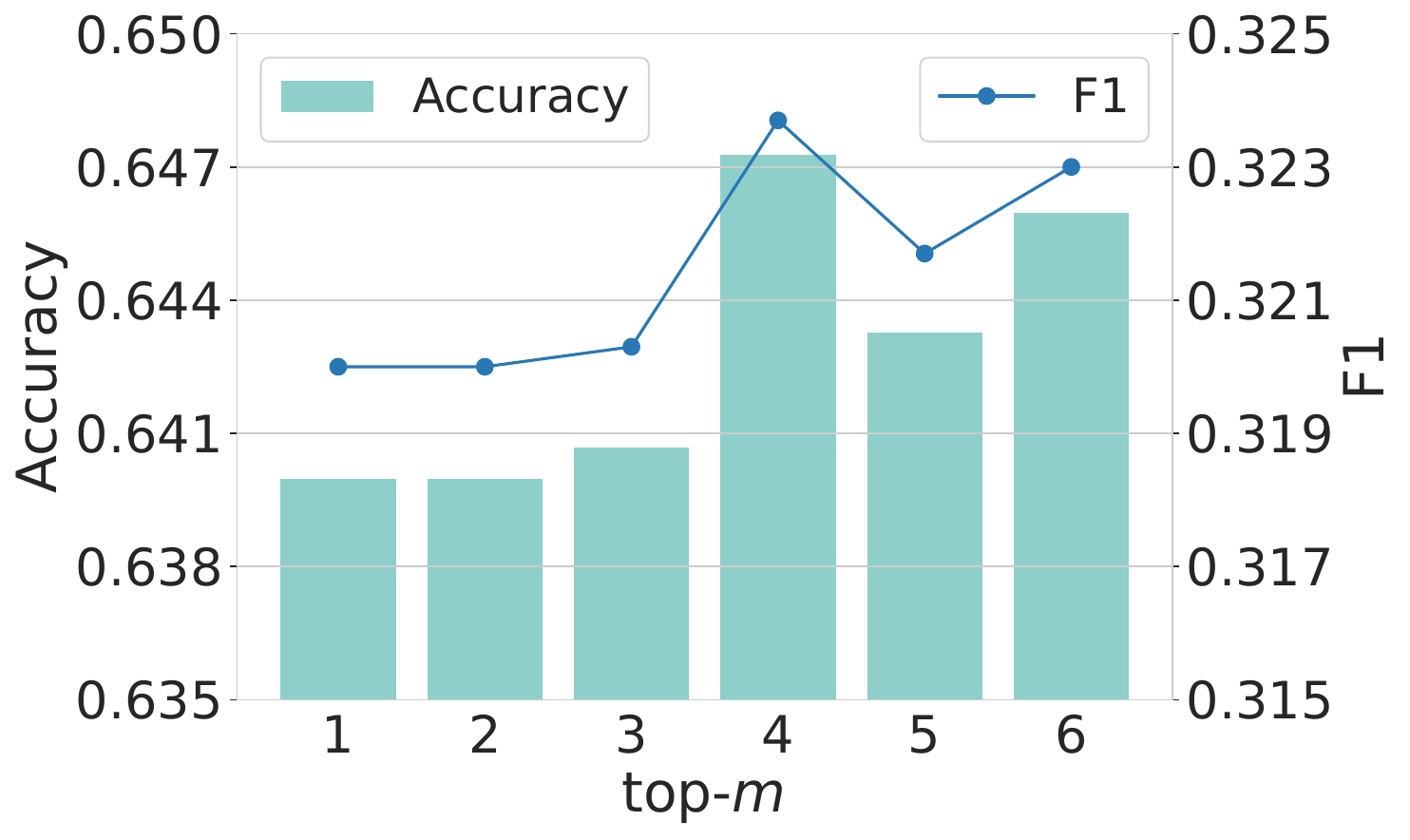}
     }
    \subfigure[LaMP-5]{
        \label{fig:n_user_lamp_5}
        \includegraphics[width=0.475\columnwidth]{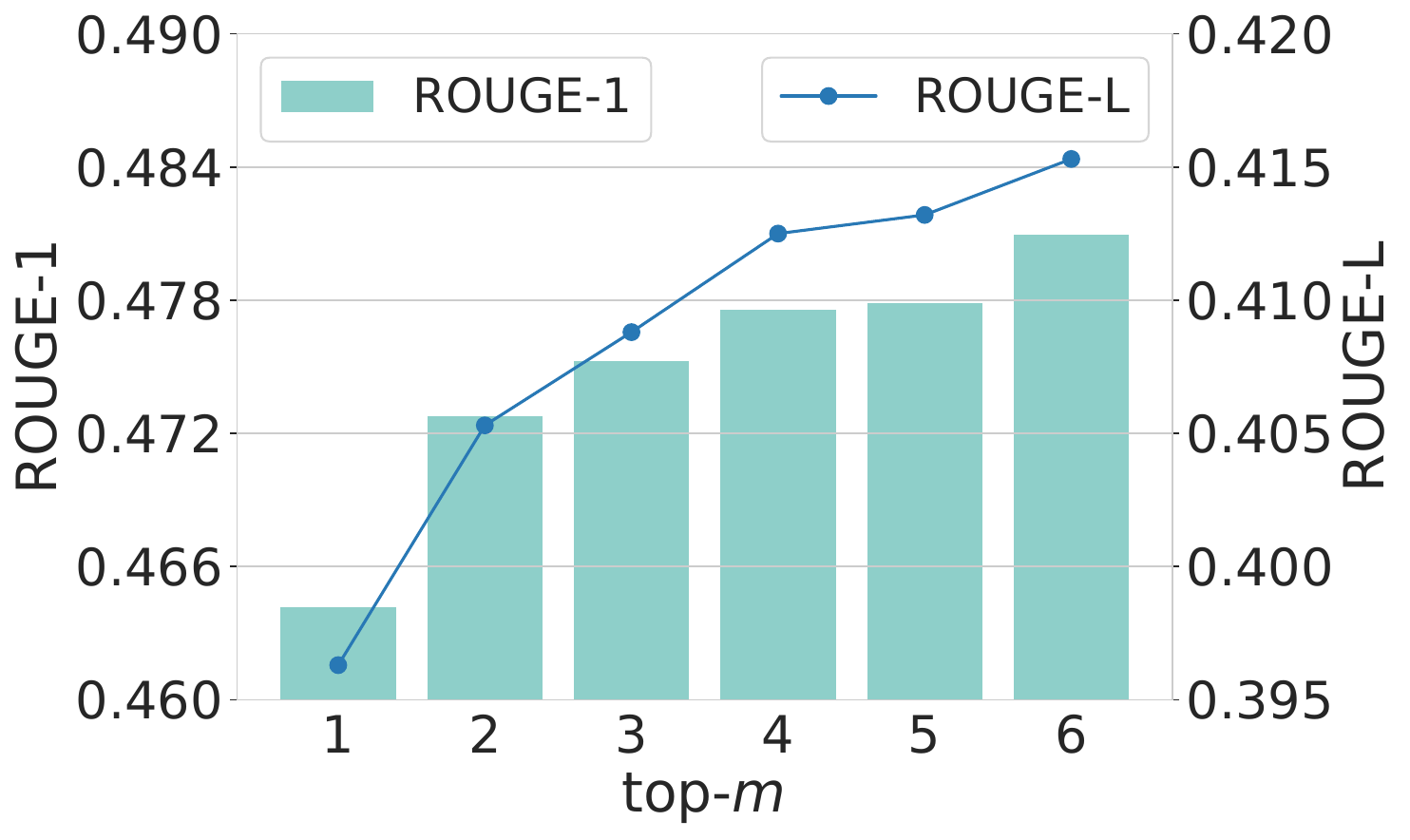}
     }
     \caption{
     Performance under different numbers of retrieved users. 
    The performance is the worst since no collaborative information is introduced when $m=1$.
     }
     \label{fig:n_user}
     \vspace{-0.3cm}
\end{figure}

\begin{figure}[t]
     \centering
     \subfigure[LaMP-1]{
        \label{fig:n_doc_lamp_1}
        \includegraphics[width=0.475\columnwidth]{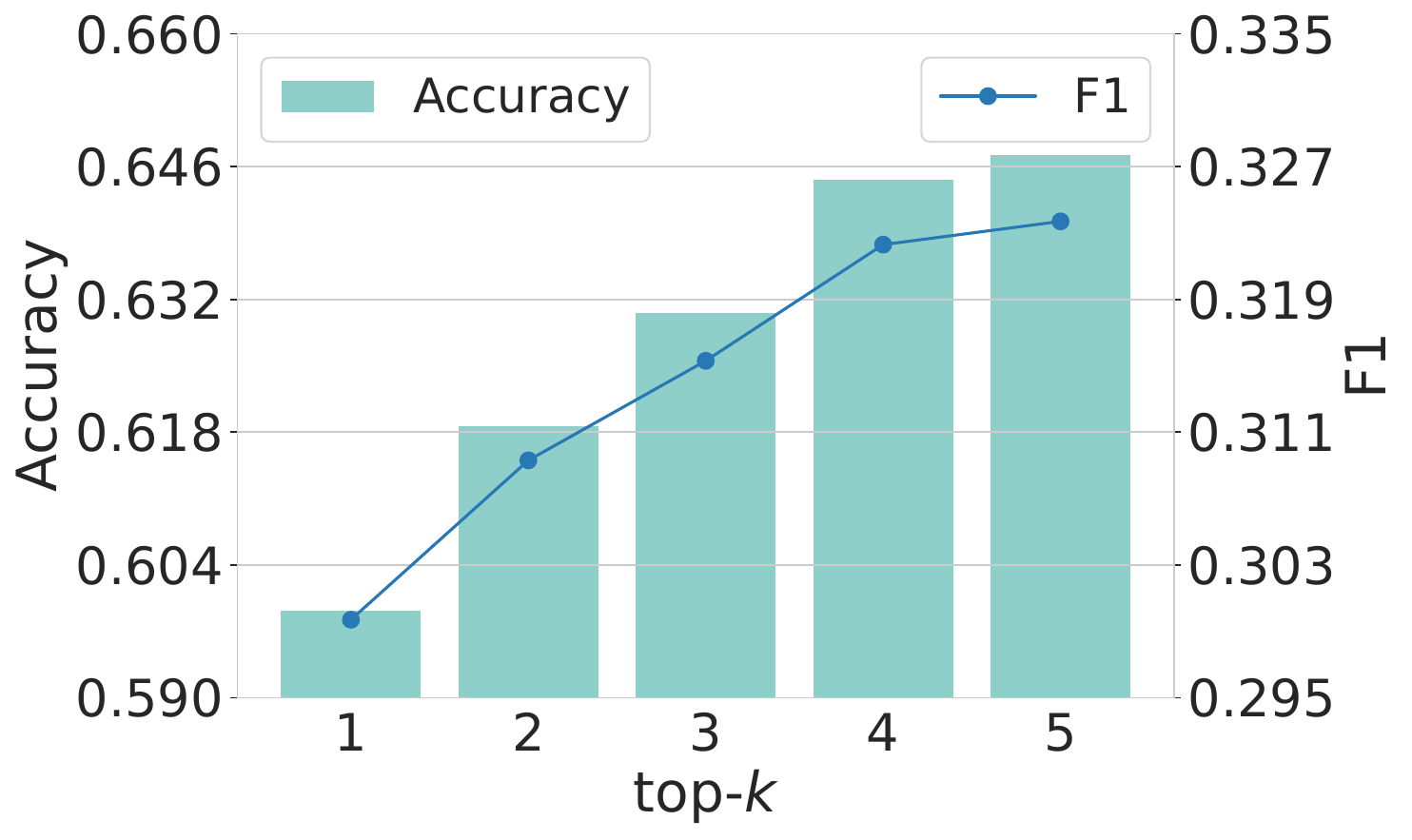}
     }
    \subfigure[LaMP-5]{
        \label{fig:n_doc_lamp_5}
        \includegraphics[width=0.475\columnwidth]{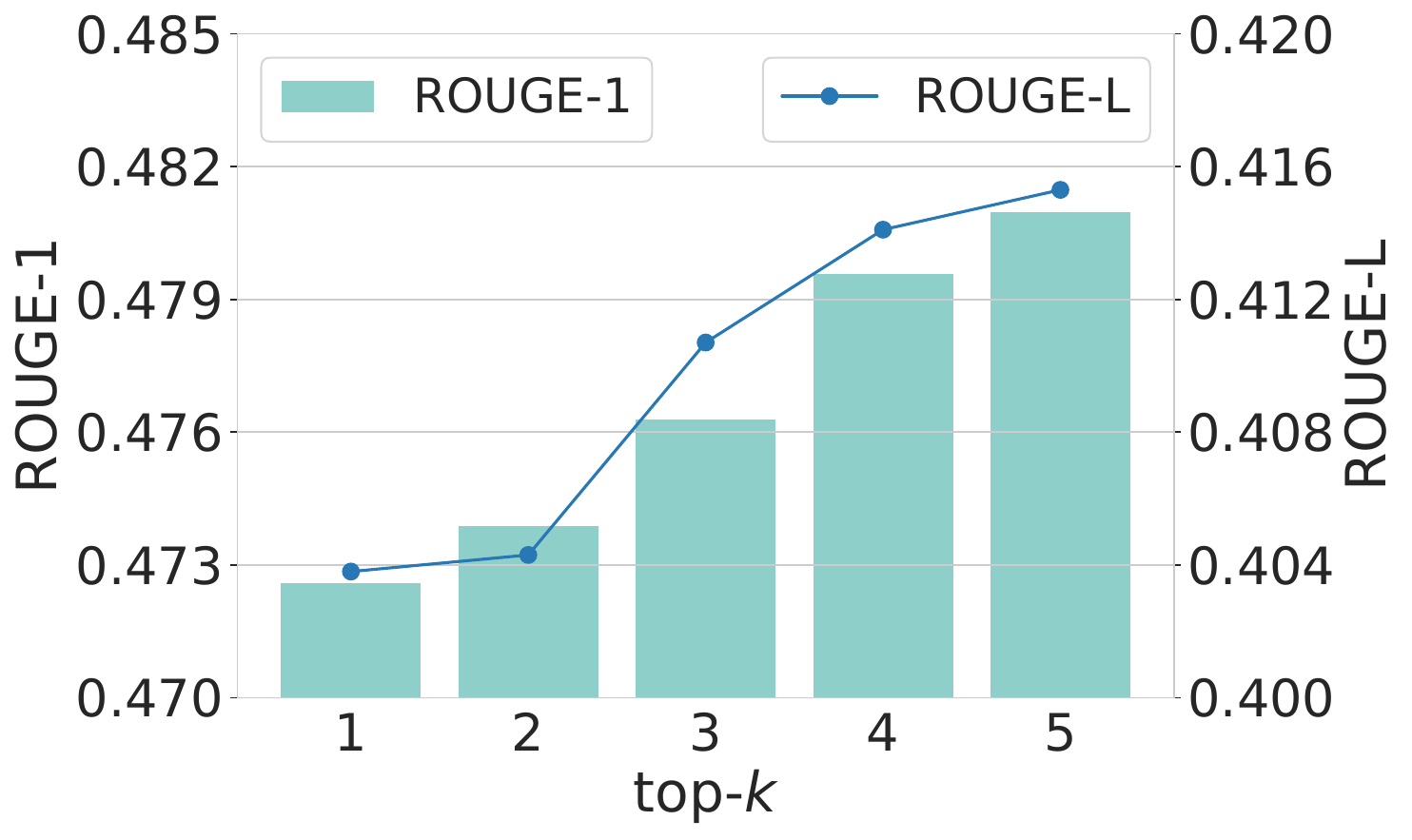}
     }
     \caption{
     Performance under different numbers of retrieved documents per user. 
     }
     \label{fig:n_doc}
     \vspace{-0.3cm}
\end{figure}

\subsubsection{Effectiveness of Document Retrieval using LLM Feedback (Challenge 2)}
As mentioned in Section~\ref{sec:intro}, to address Challenge 2, we fine-tune the retriever and reranker using feedback from the content generated by the LLM, 
enabling them to retrieve documents that better meet personalized LLM generation needs.
To validate its effectiveness, we compared the results with those using retrievers and rerankers without LLM feedback fine-tuning, as well as using BM25 as the retriever and reranker, as shown in Figure~\ref{fig:tune}. It can be observed that \ourname performs the best, highlighting the importance of fine-tuning with LLM feedback rather than relying solely on semantic relevance.

\begin{table*}[t]
\caption{The format of input, output, and user history for different datasets in the LaMP~\cite{salemi2023lamp} benchmark. In the input, \emph{\{$\text{history}_i$\}} will be replaced by the retrieved $i$-th history, and each history is represented as shown in the ``User History'' column. The other \emph{italicized text} in the input is replaced with the user's input. For text generation tasks, to ensure that the LLM does not generate irrelevant information, we instruct the LLM in the input to generate in JSON format, and then we extract the LLM's prediction from the JSON-formatted output.}
\vspace{-8px}
\centering
\resizebox{.98\linewidth}{!}{
\begin{tabular}{llll}
\toprule
\toprule
Task &Input &Output &User History \\
\midrule
LaMP-1 &\makecell[l]{The historical profiles are as follows: \emph{\{$\text{history}_1$\}} \ldots \emph{\{$\text{history}_k$\}}. \\ Based on the historical profiles provided, please choose one of \\ the following two references that is more relevant to the user's \\ input title: [1] \emph{\{$\text{reference}_1$\}}; [2] \emph{\{$\text{reference}_2$\}}. Please just answer \\ with ``[1]'' or ``[2]'' without explanation. ``title'': \emph{\{title\}}.} & [1] & \makecell[l]{``title'': \emph{\{title\}} \\ ``abstract'': \emph{\{abstract\}}}\\
\midrule
LaMP-2 &\makecell[l]{The historical profiles are as follows: \emph{\{$\text{history}_1$\}} \ldots \emph{\{$\text{history}_k$\}}. \\ Based on the historical profiles provided, please select the tag \\ from [sci-fi, based on a book, comedy \ldots] that is most relevant \\ to the user's input description. Please just answer with the tag \\ name without explanation. ``description'': \emph{\{description\}}; ``tag'': }  &comedy & \makecell[l]{``description'': \emph{\{description\}}; \\ ``tag'': \emph{\{tag\}}}\\
\midrule
LaMP-3 &\makecell[l]{The historical profiles are as follows: \emph{\{$\text{history}_1$\}} \ldots \emph{\{$\text{history}_k$\}}. \\Based on the historical profiles provided, what is the score of the \\ following review on a scale of 1 to 5? just answer with 1, 2, 3, 4, or 5 \\ without further explanation. ``review'': \emph{\{review\}}; ``score'': } &5 &\makecell[l]{``review'': \emph{\{review\}} \\ ``score'': \emph{\{score\}}}\\
\midrule
LaMP-4 &\makecell[l]{The historical profiles are as follows: \emph{\{$\text{history}_1$\}} \ldots \emph{\{$\text{history}_k$\}}. \\ Based on the historical profiles provided, please generate a title \\ for the given user's input text. Please generate it in the following \\  format: \{``title'': ``generated title''\} without explanation, and use \\ only English. ``text'': \emph{\{text\}}; ``title'': }  &\makecell[l]{\{``title'': Finding Happiness \\ After Divorce -- It Can Happen\}}  &\makecell[l]{``text'': \emph{\{text\}} \\ ``title'': \emph{\{title\}}} \\ 
\midrule
LaMP-5 &\makecell[l]{The historical profiles are as follows: \emph{\{$\text{history}_1$\}} \ldots \emph{\{$\text{history}_k$\}}. \\ Based on the historical profiles provided, please generate a title \\ for the given user's input abstract. Please generate it in the \\ following format: \{``title'': ``generated title''\} without explanation, \\ and use only English. ``abstract'': \emph{\{abstract\}}; ``title'': } &\makecell[l]{\{``title'': Link-Reliability Based \\ Two-Hop Routing for \\ Wireless Sensor Networks.\}} &\makecell[l]{``abstract'': \emph{\{abstract\}} \\ ``title'': \emph{\{title\}}}\\ 
\midrule
LaMP-7 &\makecell[l]{The historical profiles are as follows: \emph{\{$\text{history}_1$\}} \ldots \emph{\{$\text{history}_k$\}}. \\ Based on the style pattern of the historical tweets provided, \\ please paraphrase the user's input tweet without any explanation \\ before or after it. Please generate it in the following format: \\ \{``tweet'': ``generated tweet''\}  without explanation, and use only \\  English. ``tweet'': \emph{\{tweet\}}.} &\makecell[l]{\{``tweet'':\@lilxcutiesworld the \\ danny picture is GOOD!! \\ I really like it.\}} &\makecell[l]{``tweet'': \emph{\{tweet\}}} \\ 
\bottomrule
\bottomrule
\end{tabular}}
\label{tab:prompt}   
\vspace{-0.3cm}
\end{table*}

\subsubsection{Impact of the Number of Documents from the Current User}
To further validate that \ourname enhances personalization by incorporating collaborative information, we observed the impact of the number of documents from the current user in the final top-$k$ documents on the results, as shown in Figure~\ref{fig:doc_ratio}. We varied the number of documents retrieved from the current user's history in the top-$k$ documents from 0 to 5, with the remaining documents retrieved from similar users' histories. The results indicate that retrieving only from the current user's history leads to poor performance, while appropriately retrieving documents from similar users' histories significantly improves the results. This verifies the importance of incorporating collaborative information.

\subsubsection{Impact of the Number of Retrieved Users}
Since we enhance personalized text generation by introducing collaborative filtering, we further explored how much collaborative information to introduce, specifically the impact of the number of retrieved users on the results, as shown in Figure~\ref{fig:n_user}. In LaMP-1, retrieving too few or too many users leads to poorer performance, with the best results at 4 users. In LaMP-5, the performance improves as the number of users increases. This highlights the importance of introducing collaborative filtering, but it also indicates that excessive introduction can lead to decreased effectiveness.

\subsubsection{Impact of the Number of Retrieved Documents}
We also analyzed the impact of the number of retrieved documents, $k$, on the results, as shown in Figure~\ref{fig:n_doc}. It can be observed that as the number of retrieved documents increases, performance improves, indicating the importance of retrieving user history to reflect user preferences for enhancing LLM-generated results. Since more documents lead to longer prompts and slower LLM generation, we chose $k=5$ for our experiments.

\section{Conclusion}
In this paper, we propose \ourname, which adapts collaborative filtering into RAG to personalize LLMs.
To introduce collaborative information without explicit user labels and retrieve documents that support personalized LLM generation, we first train user embeddings through contrastive learning to retrieve similar users. Then, we design the personalized retriever and reranker that considers user preferences during retrieval and fine-tune them using LLM feedback.
The results on the Language Model Personalization (LaMP) benchmark validate the effectiveness of \ourname. The experimental analysis also confirms the effectiveness of each module within \ourname.

\appendix

\section{Appendix: Prompts}
\label{appendix:prompt}
We provide detailed formats for the inputs, outputs, and user histories for the LLM across different datasets, as shown in Table~\ref{tab:prompt}.

\bibliographystyle{ACM-Reference-Format}
\balance
\bibliography{ref}

\end{document}